\documentstyle[amstex,12pt]{article}                     %% AMS-LaTeX

\newcommand{\subsect}{\bf}
\newcommand{\sectwo}{{Section\,2}}
\newcommand{\secthr}{{Section\,3}}
\newcommand{\secfou}{{Section\,4}}
\newcommand{\secfiv}{{Section\,5}}
\newcommand{\secfouandfiv}{{Sections\,4 and 5}}
\newcommand{\secsix}{{Section\,6}}
\newcommand{\secsev}{{Section\,7}}
\newcommand{\seceig}{{Section\,8}}
\newcommand{\RR}{{\Bbb R}}                      %real numbers
\newcommand{\bfx}{\mbox{{\boldmath $x$}}}        %spatial x
 
\newcommand{\bfj}{\mbox{{\boldmath $j$}}} 
\newcommand{\bfp}{\mbox{{\boldmath $p$}}}        %spatial p
\newcommand{\bpi}{\mbox{{\boldmath $\pi$}}}      %representation

\newcommand{\vac}{{\Omega}}

\newcommand{\RO}{{\cal R}({\cal O})}
\newcommand{\FO}{{\cal F}({\cal O})}
\newcommand{\SO}{{\Sigma}({\cal O})}
\newcommand{\fA}{{\mathfrak A}}

\newcommand{\sA}{{\underline{A}_{}}}
\newcommand{\sB}{{\underline{B}_{}}}
\newcommand{\sC}{{\underline{C}_{}}}
\newcommand{\sfA}{\underline{{\mathfrak A}}}
\newcommand{\sa}{\underline{\alpha}}
\newlength{\dinwidth}
\newlength{\dinmargin}
\begin{document}
%%%%%%%%%%%%%%%%%%%%%%%%%%%%%%%%%%%%%%%%%%%%%%%%%%%%%%%%%%%%%
\title{\sc The Quest for Understanding \\ 
in Relativistic Quantum Physics
\thanks{Invited 
contribution to the Special Issue 2000 of the Journal of Mathematical
Physics}
}
\author{{\sc Detlev Buchholz}$\,^1$ \ and \ 
{\sc Rudolf Haag}$\,^2$ \\
$^1\,${\normalsize  Institut f\"ur Theoretische Physik,}\\
{\normalsize Universit\"at G\"ottingen,}\\
{\normalsize  D--37073 G\"ottingen, Germany}\\[10pt]
$^2\,${\normalsize  Waldschmidtstra{\ss}e 4\,b,}\\
{\normalsize D--83727 Schliersee--Neuhaus,}\\
{\normalsize  Germany}}
{\date{\normalsize \sc October 1999}}
\maketitle
${}$\\[5pt]
%%%%%%%%%Abschnittw. Nummerierung der Gleichungen%%%%%%%%%%%%
\renewcommand{\theequation}{\thesection.\arabic{equation}}
%%%%%%%%%%%%%%%%%%%%%%%%%%%%%%%%%%%%%%%%%%%%%%%%%%%%%%%%%%%%%
%%%%%%%%%Abschnittw. Nummerierung der Theoreme etc.%%%%%%%%%%
\newtheorem{Definition}{Definition}[section]
\newtheorem{Theorem}[Definition]{Theorem}
\newtheorem{Proposition}[Definition]{Proposition}
\newtheorem{Lemma}[Definition]{Lemma}
\newtheorem{Corollary}[Definition]{Corollary}
%%%%%%%%%%%%%%%%%%%%%%%%%%%%%%%%%%%%%%%%%%%%%%%%%%%%%%%%%%%%%%%%%%%%%%
%%%%%%%%%%%%%%%%%%%%%%%%%%%%%%%%%%%%%%%%%%%%%%%%%%%%%%%%%%%%%%%%%%%%%%
\begin{center} {\sc Abstract} \\
{ We discuss the status and some perspectives of 
relativistic quantum physics.}
\end{center} 
%%%%%%%%%%%%%%%%%%%%%%%%%%%%%%%%%%%%%%%%%%%%%%%
\section{Introduction}
\setcounter{equation}{0} 

The end of the first half of the century coincided with a notable
incision in the search for fundamental laws. The
breakthrough in the handling of Quantum Electrodynamics had shown that
old equations contained much more physically relevant information than
one had dared to believe. It had restored faith in the power of
quantum field theory. 
But side by side with the dominant feeling of
great triumph there was a spectrum of mixed feelings ranging from
bewilderment to severe criticism. 

Dirac emphasized that there was no
acceptable physical theory but only an ugly set of rules. Heisenberg
felt that the success of renormalization had turned the minds away
from the really important issues in shaping a new theory. Still there
was the empirical fact that QED was capable of producing numbers
agreeing with experiments to an unbelievable degree of accuracy
without any radical changes in its foundations and that there lacked any
indication that the general scheme of quantum field theory was at odds
with experiments in high energy physics, though there were obviously
great difficulties in eliminating conceptual and mathematical muddles
abounding in the existing formulation. So it appeared that the time
called for a period of consolidation, of patient work devoted to the
separation of golden nuggets from the mud. What constitutes a quantum
field theory? What is needed to extract the relevant physical
information?

It is not our intention to present 
in this essay a retrospective of developments in
the past fifty years. But it is important to recall some attitudes and
prejudices prevailing at various periods, to recall the questions
asked then and see to what extent they have been answered in a
satisfactory way in order to have a basis for the assessment of open
questions today, to recognize tasks and perspectives. Therefore we
shall begin with a brief sketch of endeavors in the fifties and
sixties. Since our article necessitated a severe restriction in 
the topics addressed and thus an unavoidable bias in the 
selection of references, it should not be used as a source for 
the ``history of science''. We shall not be concerned with 
the disentangling of ``who contributed what and when''. We shall
also suppress technical details as much as possible and refer 
the reader to the easily accessible books, where detailed 
references may be found and the methods and techniques alluded 
to are fully described. For the first sections most of this 
is given in references \cite{StWi} to \cite{BoLoTo}. 

For the most part, we shall use the language of an approach 
which is often, but inappropriately, called ``Algebraic
Quantum Field Theory'', because we feel that it provides the 
simplest and most natural formulation in which the relevant 
principles can be expressed and it also provides a powerful
mathematical structure which can be precisely described 
and applies to a wide area.
No quantum fields appear in this formulation. 
In fact, the relation to quantum fields is not as close as originally believed.
In particular, it is important that 
it can also incorporate extended objects which generalize the 
field concept. Thus a better name is ``Local Quantum Physics''.
For details and references see \cite{Ha}. 

Our main aim is to describe the questions that presented themselves
at various times, follow the changes of perspective needed in 
answering them and indicate open questions to which we do not know 
the answer and which might suggest tasks to think about in the 
future.  
%%%%%%%%%%%%%%%%%%%%%%%%%%%%%%%%%%%%%%%%%%%%%%%%%%%%%%%%%%%%%%%%%%%%%%
%%%%%%%%%%%%%%%%%%%%%%%%%%%%%%%%%%%%%%%%%%%%%%%%%%%%%%%%%%%%%%%%%%%%%%
\section{Taking stock}
\setcounter{equation}{0}
In the construction of models in quantum field theory one usually
starts from a classical field theory and tries to ``quantize'' it
following as closely as possible the rules which had proven so
successful in the transition from classical to quantum
mechanics. The
dynamical variables are now a set of fields which transform
covariantly under some finite dimensional representation of the Lorentz
group (e.g.\ spinors, vectors). The key element in characterizing the
model is the Lagrangian from which the equations of motion and
commutation relations can be guessed. One novel feature appeared: For
some fields the commutator had to be replaced by the anticommutator 
in order to comply with the Pauli principle. 

This scheme was
immediately successful in the case of free fields. Such a field can be
decomposed into a positive and a negative frequency part, yielding
annihilation and creation operators for some type of particle. The
theory then just describes an arbitrary number of identical, non--interacting
particles. This feature was interpreted as a manifestation of the well
known wave--particle dualism. There was, however, no easy way to extend
this formalism to a theory of interacting fields. It became clear that
the commutation relations could no longer be the canonical ones but
must have stronger singularities; that the equations of motion were
sick because the product of fields at the same point defied any
simple definition; that one had to think more carefully about the 
relation between fields and particles. 

What to keep and what to throw overboard? We shall divide the
tentative answers given into two groups. The first, 
group $A$,  concerns the
general setting, the second, 
group $B$, the field concept. In the first group we
have: \\[2mm]
  { $A_1$) \ Keep the notion of space--time as a classical manifold with
  pregiven geometry (the Minkowski space $\cal M$) as the arena in
  which physics plays. Its symmetry group is the ``Poincar\'e group'',
  generated by translations and Lorentz transformations.\\[2mm]
  $A_2$) \ Keep the standard formalism of quantum physics in which pure
  states are described as ``rays'' in a Hilbert space $\cal H$ (unit
  vectors up to a phase factor) and observables as self adjoint
  operators acting in $\cal H$.\\[2mm]
  $A_3$) \ Incorporate the results of Wigner's analysis: A symmetry is
  implemented by a ``ray representation'' of the symmetry group. In
  the case of the Poincar\'e group $\cal P$ this is equivalent to a
  representation of the covering group $\widetilde{\cal P}$ by unitary
  operators. \\[2mm]
\indent  This provides already important physical information. For
  instance, the infinitesimal generators of the translations $
  P^\mu$ may be interpreted as observables 
  corresponding to the total energy--momentum. It is a purely
  mathematical problem to determine all irreducible representations
  and this problem has been solved. It turns out that an irreducible
  representation with positive energy describes the state space of a
  single stable particle. All other representations can be constructed
  by direct sums and tensor products from the irreducibles.}  
Since the restriction to positive energies seems to be well motivated
(for instance to ensure stability), one comes to the first basic
postulate (axiom, principle): \\[2mm]
{ $S$) \  The spectrum of the energy--momentum operators $P^\mu$
in $\cal H$ is restricted to the closed forward cone 
$\overline{V}_+ = \{ p : p_0 \geq |\bfp | \}$. 
One usually also assumes that there is a unique ground state 
$\vac$, the vacuum.} \\[2mm]
\indent 
Since we are talking about field theory, we decide in group $B$:\\[2mm]
{ $B_1$) \  Keep the idea that the basic dynamical variables, in terms
of which all operators in $\cal H$ should be expressed, are fields.} \\[2mm]
\noindent 
The naive idea that a field $\varphi$ assigns to each space--time
point $x$ an operator $\varphi (x)$ in $\cal H$ is not tenable.
Therefore a
considerable amount of mathematical care and sophistication is needed
to avoid pitfalls. One may consider a field as an
 ``operator valued distribution'' on a suitably defined domain in $\cal
H$ or as a sesquilinear form on this domain. This being done, one may
formulate the postulate $B_2$:\\[2mm]
$B_2$) \  The theory is completely described by a finite number of covariant
fields (each having a finite number of components). \\[2mm]
\indent
Why fields? This question is asked at regular intervals. One strong
argument was, of course, the success in QED. But more deeply, the
notion of field allows us to encode the relativistic causal structure
of space--time in the theory and this is implemented by the basic
locality postulate.\\[2mm]
$L$) \ Field quantities in regions which lie space--like to 
each other either commute or
anticommute. \\[2mm]
\indent
Experiments in high energy physics are concerned with particles and
cross sections, not with fields. So one needs to know the connection
between fields and particles. An important step in this direction were
the asymptotic relations of Lehmann, Symanzik and Zimmermann 
\cite{LeSyZi} which
provided an elegant algorithm relating correlation functions of fields
to S--matrix elements.

The strategy of starting from precisely defined postulates, analyzing
their consequences and focusing first on general structure instead of
specific equations, created an enterprise with rather novel style
(``Axiomatic Quantum Field Theory''). The emphasis on mathematical
rigor, stating results in the form of theorems and lemmas, was
instrumental in establishing a very fruitful discussion between
mathematical \mbox{physicists} and pure mathematicians, closing a deplorable
gap. On the other hand, it was not to the taste of all parts of the
physics community, as illustrated by a joke circulated in the early
sixties: ``The contribution of axiomatic quantum field theory to
physics is smaller than any preassigned positive number
$\varepsilon$''. To balance this joke, we should also mention
another one: ``In the thirties, under the demoralizing influence
of quantum theoretic perturbation theory, the mathematics required
of a theoretical physicist was reduced to a rudimentary knowledge
of the Latin and Greek alphabets'' (Res Jost, as quoted in 
\cite{StWi}). 

Well, there is no point in arguing with good jokes. 
But they do contain messages which should be taken seriously.
A belief that mathematics is the prime mover for 
progress in physics is not warranted and if it leads to an 
overemphasis on mathematical rigor, there is the 
danger of distracting attention from the essential points
and contributing to a language problem so that different camps
abandon the effort to understand each other's vocabulary.
On the other hand, a narrow
view of what constitutes ``real physics'' fosters an ill founded 
snobbism. It takes many kinds of craftsmen to construct a building.
The enterprise whose origins were sketched above did contribute 
to ``real physics'' in many ways. 
Not only by clarifying issues and proving or disproving 
conjectures but also by providing tools, essential for many 
subsequent developments. A prime example are the 
analyticity properties of $n$--point functions, derived as
consequences of postulates $S$ and $L$ by Wightman \cite{Wig},
seminal for a variety of subsequent developments
(dispersion relations, renormalization theory, Euclidean 
formulation etc.).  
And it raised new questions. One of them concerned a deeper
understanding of the relation between fields and particles.
We shall devote the next section to this. 

%%%%%%%%%%%%%%%%%%%%%%%%%%%%%%%%%%%%%%%%%%%%%%%%%%%%%%%%%%%%%%%%%%%%%%
%%%%%%%%%%%%%%%%%%%%%%%%%%%%%%%%%%%%%%%%%%%%%%%%%%%%%%%%%%%%%%%%%%%%%%
\section{Fields and particles}
\setcounter{equation}{0}
In the early years of quantum field theory, the prevailing picture was:
There are a few types of ``elementary'' particles which serve as the
building stones for more complex structures (from nuclei to crystals)
and to each species of elementary particle there corresponds a
fundamental quantum field. If this is a good picture then the
relation between a basic field and the states of the corresponding
particle (isolated before or after a collision) is adequately
described by the LSZ--formalism and these relations may be regarded as
an ``asymptotic condition'' needed for a reasonable interpretation of
the theory. The problem of how to deal with composite particles
(``bound states'') could be postponed as a later worry.

But evidence from a variety of sources eroded this simple picture. It
was difficult to decide whether newly discovered particles should be
regarded as elementary and honored by associating a new basic field
with them. It was recognized that no simple and clear distinction
between elementary and composite was available, neither for fields nor
for particles. In the process of developing a collision theory for
composite particles, it was discovered that the LSZ--formalism could be
rather easily extended to cover also this case. Furthermore, there was
no need for any close connection between particle type and basic
field. The asymptotic condition was not an extra condition but a
consequence of the postulates $S$ and $L$. All that was needed was the
existence of a discrete part in the mass spectrum (single particle
states, no matter whether elementary or composite) and the existence of
some ``quasilocal operators'' connecting the vacuum with these
states.\footnote{For a long time (in many textbooks till present
  days) the formulation of collision theory was based on the
  comparison of a ``free Hamiltonian'' $H_0$ with the full Hamiltonian
  $H$. To adapt this formulation to many channel reactions with
  ``bound states'', even in non relativistic quantum mechanics one had
  to find a different separation of free motion and interaction for
  each channel, a procedure which was
  not only difficult but highly nontransparent if due account of the
  Pauli principle was taken (the dispute about ``post-prior
  antisymmetrization''). To our knowledge, the first natural approach
  to the problem was due to H. Ekstein \cite{Ek} (compare also 
  \cite{Ha1,BrHa}). The results in quantum field theory mentioned
  above were obtained in full clarity by D. Ruelle \cite{Ru}.} 

The term ``quasilocal'' brings us back to the 
original significance of the field concept, namely the establishment
of a relation between space--time and the dynamical variables of the
theory, to allow us to characterize those operations which (at least
approximately) pertain to a specific region in space--time.
In other words: there is no general field--particle duality. A particle
is a stable quasilocal excitation. The determination of the types of
particles appearing in the theory is a dynamical problem which bears
some analogy with the determination of the ground state of an atom
with one important difference. We cannot regard the particle as a
composite of discrete, elementary objects.

But what about leptons, quarks and the parton picture? Is the success
of the ``Standard Model'' not evidence to the contrary? Not really. The
message of the Standard Model has much in common with the message
received from QED. It does suggest that a field theory in which the
property $B_2$ is concretized by a specific set of fermionic fields and
gauge fields has more physical relevance than one dared to hope. But
again this achievement is accompanied by a host of puzzling
questions. {}From a formal point of view we have a successful field
theory encompassing the postulates $B_2$ and $L$. But the dynamical
variables do not operate in the space of 
physical states \cite{St,KuOj}. The road to
the Hilbert space to which the above mentioned items $A_2$, $A_3$ and
$S$ refer is quite involved. This applies a forteriori to the description
of particles, as illustrated by the fragmentation models used in the
discussion of jets in high energy reactions. The important progress,
the salient feature, is the discovery of a relevant set of charge
quantum numbers (color, flavor, electric, weak). The finite number of
elementary objects refers to these, not to particles. The basic fields
are the vehicles to handle the creation and transport of such
charges. But none of these fields is observable in the sense of $A_2$.

This accentuates an old question. If postulate $L$ intends to express
only the relativistic causal structure of space--time, then it should
simply read\\[2mm]
$L_O$) \ Observables relating to space--like separated regions
commute. \\[2mm]
The intrinsic information of the theory (as contrasted to the
particular way in which the theory is described) ultimately concerns
the relation between observables. This raises the question: How can we
characterize the intrinsic structure and what is the role of quantum
fields in it? This will be addressed in the next section.

%%%%%%%%%%%%%%%%%%%%%%%%%%%%%%%%%%%%%%%%%%%%%%%%%%%%%%%%%%%%%%%%%%%%%%
%%%%%%%%%%%%%%%%%%%%%%%%%%%%%%%%%%%%%%%%%%%%%%%%%%%%%%%%%%%%%%%%%%%%%%
\section{Fields and algebras}
\setcounter{equation}{0} 

If we want to adhere to the first group of assumptions, from $A_1$ to $S$
in \sectwo, and want to incorporate the causality principle in the
form $L_O$, then we must classify the observables according to
space--time regions, i.e.\ focus on a correspondence
\begin{equation}
\cal O \longrightarrow \RO
\end{equation}
between space--time regions $\cal O$ and the algebras $\RO$
generated by the observables in the respective region. 
More precisely,
${\cal O}$ shall denote an open, bounded region
in $\cal M$.\footnote{To fix ideas, one may think of ``double--cones'',
\label{dc} the causal completions of open balls in space,  
e.g.\ ${\cal O}_r = \{x : |x_0| + |\bfx| < r \}$, and their Lorentz
transforms.} As elements of
the algebras we may take bounded operators acting in $\cal{H}$,
thereby escaping the complications with domain problems. After all, the
most elementary observables are projectors. This suggests that we specify
$\RO$ to be von Neumann algebras 
(also called von Neumann rings).\footnote{A brief account 
\label{math} of some basic notions in the theory of operator algebras is given
in the Appendix. For a thorough understanding of
this area of mathematics see, for example, \cite{KaRi,BrRo}.} 
It appears to be
the most natural choice since a von Neumann--algebra is the set of all
bounded operators which commute with a given set of others. This fits
well with the form $L_O$ of the causality requirement. 

In addition to the von Neumann algebras $\RO$, acting
on $\cal H$, we have a representation of the Poincar\'e group
by unitary operators, satisfying the spectrum condition $S$, and 
a distinguished Poincar\'e invariant pure ground state, the vacuum.
An indiv\-idual algebra $\RO$ carries hardly any physical information.
It is the relation between algebras of different regions and the 
correspondence (4.1) which contains the physics. This correspondence
requires, on the mathematical side, that the set 
$\{ \RO \}$ of algebras 
satisfies certain structural requirements \cite{Ha}.
The essential ones are 
\begin{itemize}
  \item[a)] inclusion relations (``isotony''), i.e.\
$ {\cal O}_1 \subset {\cal O}_2 \ \mbox{ implies} \
{\cal R} ({\cal O}_1) \subset 
{\cal R} ({\cal O}_2), $
\item[b)] causality (the principle $L_O$), and
\item[c)] covariant action of the Poincar\'e group on the local
  algebras. 
\end{itemize}
We shall call the set  $\{ \RO \}$ of algebras, labelled by the 
space--time regions $\cal O$ and endowed with this structure, the
``net'' of local algebras. 

It is clear that the correspondence (4.1) provides a starting 
point for the physical interpretation. It is, however, remarkable
that nothing more is needed. In other words, the net of algebras 
defines the theory, including the full physical interpretation. 
Once the net is given, we can analyze its physical predictions in 
terms of particles, collision cross sections, etc. We shall not
describe this here but just indicate the reasons. The net allows us 
to construct the mathematical counterparts of coincidence
arrangements of detectors. The specification of what the
individual detector detects need not be fed in. 
It suffices that we can extract information about the energy--momentum
range it selects, using the action of the Poincar\'e group, and 
that we have information about the placement, using the net structure. 
Exploiting judiciously
just the information from different geometric constellations one is
able to disentangle the particles and their collision cross sections
which are described by the net. 
In this analysis the spectrum condition $S$ plays a
significant role. The essential arguments are given in 
\cite{ArHa,BuPoSt}. For further details see \cite{Ha}.

Thus we may say that a net 
$\cal O \longrightarrow \RO$ gives an intrinsic
description in which the physically relevant information is
encoded. One might therefore conjecture that quantum fields should be
regarded as a (more or less convenient) way to coordinatize the
net. This point of view is supported by a result which Borchers
obtained in the context of the LSZ formalism \cite{Bo}. We get, for
instance, the same physical information whether we consider a free
field $\varphi_0$ or its Wick power $\varphi = \, : \! \varphi_0^3 \! :$ 
as the basic
field, though there is an obvious difference in convenience. A less
trivial example, where the identification of two at first sight very
different looking field theories required much more work can be found
in \cite{Re}. Other surprising examples have attracted much
attention. So the equivalence of the Thirring model with the
Sine--Gordon model \cite{Co} and recently explored equivalences in
supersymmetric Yang--Mills theories \cite{Se}.

As experienced in other areas of study, depending on the problem and
on the taste of the investigator, there are advantages in the use of
coordinates and there are advantages in using an intrinsic (coordinate
free) formulation. So the clarification of the relation between fields
and algebras is an important issue.\\[2mm]
{\subsect {}From fields to algebras:}\\[2mm]
Heuristically one would like to define $\RO$ as the von Neumann algebra
generated by the (smoothed out) observable fields in the region $\cal
O$. Appealing to von Neumann's double commutant theorem \cite{KaRi}, 
this may be symbolically written as\footnote{The commutant 
${\cal S}^\prime$ of 
a set ${\cal S}$ of operators consists of all bounded operators commuting
with the elements of ${\cal S}$; the double commutant ${\cal
  S}^{\prime \prime}$ is the commutant of ${\cal S}^\prime$.
It is assumed that the sets contain with every operator also its
adjoint. This must be required correspondingly for the set of 
fields in (4.2).} 
\begin{equation}
\RO = \{ \varphi (x) : x \in \cal O \}^{\prime \prime}, 
\end{equation}
where $\varphi$ stands for the set of observable fields. Because of
subtle questions concerning the commutativity of unbounded operators, it
is, however, not clear from the outset whether this heuristic idea can
be really implemented. The first steps in the analysis of this problem
were taken by Borchers and Zimmermann \cite{BoZi}. They showed that if
the vacuum $\vac$ is an analytic vector for the fields, i.e. if the
formal power series of the exponential function of a smeared field
applied to $\vac$ converges absolutely, then the passage from fields to
local algebras via (4.2) can be accomplished. Further progress on this
problem was made in \cite{DrFr}, where it was shown that fields
satisfying so--called linear energy bounds generate acceptable nets of
local algebras. This result covers most of the interacting quantum field
theories which have been rigorously constructed so far in the
endeavor of ``constructive quantum field theory''. 

As for the general
situation, the most comprehensive results are contained in \cite{BoYn}
and the references quoted there. In that analysis certain specific
positivity properties of Wightman functions were isolated as the
crucial prerequisite for the passage from fields to
algebras. Altogether the result of these investigations could be
summarized by saying that, while the original form of the Wightman
axioms is not sufficient to allow the transition from fields to algebras
along the lines indicated by (4.2), this can be remedied by adding
some rather unsuspicious requirements. 

More serious is the fact that in (4.2) we were talking about
``observable fields''. As already indicated, the development of field
theory has led to a situation in which none of the basic fields is
observable. The proper assessment of this problem becomes, however,
clearer in following the opposite road. \\[2mm]
{\noindent \subsect {}From algebras to fields:} \\[2mm]
As already mentioned, the characterization of the theory by a net of
local algebras is more general than the traditional field theoretic
approach. It covers also the case of observables which are not built
from point--like objects but are localizable in extended (though finite)
regions, such as Wilson loops or (finitely extended) Mandelstam
strings. Nevertheless, the point field content is of great interest
since we believe that it contains such distinguished observables as
the components of the energy momentum tensor, certain currents
etc. Heuristically, the point fields can be recovered from the algebra
by a formula like
\begin{equation}
\{\varphi(x)\} = \bigcap_{{\cal O} \, \ni \ x} \overline{\RO}. 
\end{equation}
The bar on the right hand side indicates that one cannot take the
intersection of the local algebras themselves, which is known to
consist only of multiples of the identity. Therefore, one first has to
complete the algebras in a suitable topology which allows the 
appearance of unbounded operators, respectively linear forms. This was
carried through in \cite{FrHe}, where the needed completion of the
local algebras was defined with the help of ``energy norms'' which are
sensitive to the energy--momentum transfer of the observables. Using
this device, it was shown that, provided that the algebras are
generated from sufficiently regular fields in the sense indicated by
(4.2), one can recover the fields from the algebras via (4.3). {}From a
general point of view it would be desirable to clarify the status of
point fields without assuming their existence from the outset. An
interesting proposal in this direction was recently made in
\cite{HaOj}. We shall come back to it in Section 8. \\[2mm]
{\noindent \subsect Unobservable Fields:} \\[2mm]
One circle of nagging questions was known for a long time but mostly
regarded as of minor importance. It begins with the original
formulation $L$ of the causality principle. Why the Bose--Fermi
alternative? In the sequel of his discussion of the ray representations
of the Poincar\'e group, Wigner noted that the relative phase between a
state vector belonging to a double valued representation (spinorial
wave function) and one belonging to integer spin could have no
physical meaning. Then it was recognized that such limitations of the
superposition principle occur also between states of different
electric charge due to the principle of gauge invariance in QED and
that they may be expected in still other circumstances \cite{WiWiWi}.
This called for a modification of assumption $A_2$ in \sectwo: The
Hilbert space decomposes into a direct sum of mutually orthogonal
subspaces, called the ``coherent sectors'', and the relative phase
between state vectors in different sectors is void of physical
meaning. The unobservable fields can then be regarded as operators
leading from one sector to another. One might be inclined to accept
these so--called ``superselection rules'' as a fact of life, producing
some slight complication such as the appearance of unobservable
fields. But this is somewhat artificial and
calls for a more natural explanation.

Could it be that the coherent sectors were just the modules 
(representation spaces) of
inequivalent representations of one basic algebra? Let us remember
that, at the birth of quantum mechanics, Dirac introduced the notion of
``q--numbers'' defining some abstract algebraic structure and that the
equivalence of this with the wave mechanical formulation (i.e.\ 
ultimately with Hilbert space operators) depended on the uniqueness
proof for the representation of the canonical commutation
relations. Now it was known (in circles of mathematical physicists
since the early fifties) that in the case of infinitely many degrees
of freedom the uniqueness theorem failed. In fact, there was an 
innumerable host of inequivalent representations of canonical
commutation relations. So it seemed that for the
interpretation of the theory one needed more than an abstract
algebra. 

In mathematics
the theory of a class of abstract algebras which allowed
representations in Hilbert space, the so--called 
``C*--algebras'', had
been developed. Therefore Irving Segal, one of the fathers of the
mathematical theory, had advocated for several years to
base the physical theory on an abstract C*--algebra. But quantum field
theorists who were aware of the difficulty of an appropriate physical
interpretation and of the problem of overabundance of inequivalent
representations had no use for this advice. 

Two things were necessary
before the idea of using abstract algebras could be implemented. On
the one hand, strange as it may seem in retrospect, one had to
recognize that we are not talking about a single algebra but about a
net of algebras 
whose interpretation was hinged to space--time. Secondly, one had to
realize that unitary inequivalence of representations was a much too
fine distinction to be of any physical relevance because we can
measure only with finite accuracy and consider only a finite number of
observables at a time. Thus it was indeed possible and reasonable to
consider the abstract algebraic structure as the primary definition of
the theory and Hilbert space and representations as secondary. 

To avoid confusion of concepts, we shall in the following use the Gothic
letter $\fA$ for a C*--algebra and $\cal R$ for a von
Neumann algebra, the symbol $\bpi$ to
denote a representation$^{\, {\ref{math}}}$. Thus $\bpi (\fA )$ is a 
concrete algebra of operators in a Hilbert space. 
It leads us back to a von Neumann algebra, the double commutant
\begin{equation}
\cal R = \bpi (\fA)^{\prime \prime}.
\end{equation}
The reformulation of the theory so that the local algebras 
are considered as abstract C*--algebras $\fA ({\cal O})$ 
was done in \cite{HaKa}. 

So one had reached a point in relativistic quantum physics, reminiscent
of the situation in quantum mechanics in 1926, where the primacy of
algebraic relations was emphasized as the essence of the
theory. It did suggest a natural way to understand the appearance 
of different coherent sectors. But it raised new questions. On the one hand
it was apparent that the selection of those representations usually
discussed in quantum field theory resulted from some convenient
idealizations, especially from simplifying assumptions concerning the
physical situation at space--like infinity. A closer look at the
``states of physical interest'' showed, however, that this was not the
whole truth. We have more information. 
This will be discussed in the next section.

%%%%%%%%%%%%%%%%%%%%%%%%%%%%%%%%%%%%%%%%%%%%%%%%%%%%%%%%%%%%%%%%%%%%%%
%%%%%%%%%%%%%%%%%%%%%%%%%%%%%%%%%%%%%%%%%%%%%%%%%%%%%%%%%%%%%%%%%%%%%%
\section{States of physical interest}
\setcounter{equation}{0}

In the algebraic setting a ``state'' $\omega$ 
is considered as a positive, linear and normalized 
functional over the algebra $\fA$. It assigns 
to each $A \in \fA$ a complex number. It is real for
self--adjoint elements and is then  
interpreted as an expectation value. Any such state  
gives rise to a representation $\bpi$ of the algebra
on some Hilbert space $\cal H$, where it can be
described by a unit vector (GNS--construction).$^{{\ref{math}}}$ 
Convex combinations of states give again states. This 
mathematical operation corresponds to the physical 
procedure of ``mixing''. Conversely, pure
states are the extremals in a convex decomposition 
(which, sometimes, may be physically meaningless).  

In physics we consider primarily two classes of states, corresponding
to different situations. In ``particle physics'' we are interested
in states which are close to the vacuum, differing from it only by
some more or less localizable disturbances. In statistical mechanics
we are interested in states which are close to a thermal 
equilibrium state. These are idealized best by considering a medium 
with non--vanishing density, extending to infinity.\footnote{Though
one may focus on a system in a finite volume, this does not really 
change the above statement because one then has to specify the
relation to the outside either by introducing a heat bath or by artificial
boundary conditions.}

One bonus of the formulation of the theory in terms
of local algebras is that the common features of both classes become
apparent and some powerful tools for a structure analysis in both
areas emerge. Not only has the characterization of a thermal
equilibrium state by the ``thermal boundary condition'', arising from the
work of Kubo and of Martin and Schwinger, a very
simple form in the algebraic setting \cite{HaHuWi}. But it turned
out surprisingly that this so--called KMS--condition has a very natural
place in mathematics, the Tomita--Takesaki theory of modular
automorphism groups \cite{Ta}, 
which plays a central role in the classification of
von Neumann algebras.$^{{\ref{math}}}$ 

It was a remarkable experience that the ideas of
Tomita and the paper \cite{HaHuWi} were presented at the same
workshop, motivated by entirely different purposes and in complete
ignorance of each other. This led to an intensive interaction between
some groups of mathematicians and physicists from which both sides
profited substantially. What is the crux of the matter? Suffice it
here to say that Tomita and Takesaki studied von Neumann algebras for
which there existed a vector which is both cyclic and separating 
for the algebra$^{\ref{math}}$.
They found that such a vector (or rather the corresponding
state) defines a distinguished one--parameter automorphism group for the
algebra with some remarkable properties and a conjugation 
mapping of the algebra on its commutant. This group of modular
automorphisms plays also an important 
role in physics. For instance, the extension of
Gibbs' characterization of thermal equilibrium states to an infinitely
extended medium is equivalent to the statement that equilibrium 
is described by any state whose modular group is some one--parameter 
subgroup of time translations and (global) gauge transformations.
(In the non--relativistic limit, the latter corresponds to the  
conservation laws for independent species of particles instead of 
charges.)  

In the case of zero temperature this
formalism degenerates. In particular, the vacuum vector is not
separating for the global algebra $\bpi (\fA)$ of all observables. 
However, a theorem of Reeh and
Schlieder \cite{ReSch} tells us that it is cyclic and
separating for $\cal{R(O)} = \bpi (\fA ({\cal O}))^{\prime \prime}$ 
whenever there is a non--void causal
complement of the region $\cal O$. What can we say about the modular
automorphism induced by the vacuum for such algebras? The first
important discovery in this context was made by Bisognano and Wichmann
\cite{BiWi} who determined these 
automorphisms for special regions, called ``wedges'', 
such as 
\begin{equation} \label{wedge}
W = \{ x : x_1 > |x_0|, \ x_2, x_3 \ \mbox{arbitrary} \}.
\end{equation}
They found that these automorphisms coincide with the 
Lorentz boosts, leaving the wedge invariant. 
The close connection of this fact to the Bekenstein--Hawking
temperature of black holes was recognized somewhat later 
by Sewell \cite{Sew} and more fully discussed 
in \cite{HaNaSt}. In the case of theories
with conformal invariance, such a geometric significance of modular
automorphisms could also be established for double 
cones \cite{HiLo}.\footnote{For the full development of applications of modular
theory to quantum field theory, see the contribution of 
H.--J.\ Borchers.}

Besides such specific
identifications, it was gradually realized that the von Neumann
algebras of finitely extended regions are all of one universal type,
irrespective of whether we consider thermal states or states in
particle physics and that this is a consequence of important generic
properties of the physical states. Again, this development originated
from a bunch of quite different questions. \newpage
{\noindent \subsect Phase space properties:}\\[2mm]
Since the connection between fields and particles is not very close
and we even know field theoretical models which have
no particle content whatsoever, one must ask for the conditions under
which particles appear in the theory. We stated earlier that states of
particles are quasilocal excitations. Naively, one would be inclined to
define ``localized states'' by
application of a local algebra $\cal{R(O)}$ to the vacuum. But the
Reeh--Schlieder theorem tells us that this leads to a
dense set in $\cal H$ and all reminiscence of the region $\cal O$ is
lost. The reason for this paradox is that the vacuum state
incorporates correlations between observables in far separated regions
which cannot vanish exactly because of analytic properties of the
correlation functions. 

For a vector
\begin{equation}
\Psi = A \vac \quad \mbox{with } \ A \in \RO
\end{equation}
there is the ratio between ``cost and effect'', 
\begin{equation}
c_A = {{ \parallel A \parallel} \over {\parallel \Psi \parallel}},
\end{equation}
which in general is larger than $1$. If $c_A$ is close to $1$,  
then $\Psi$ does describe an excitation which
is approximately localized in $\cal O$, i.e.\ the 
expectation value of an observable in a region space--like to 
$\cal O$ in this state is approximately equal to the
vacuum expectation value. But as $c_A$ gets larger, this
significance of $\cal O$ for the interpretation of $\Psi$ is lost. In
other words, denoting the unit ball of $\cal R$ by ${\cal R}_1$
(the set of all $A \in \cal{R}$ with $\parallel A \parallel \le 1$), 
the set of vectors 
${\cal R}({\cal O})_1 \vac$ characterizes a part of
$\cal H$ which, apart from vectors of very small length, describes
approximate localization in $\cal O$. 

If we choose for $\cal O$ a
bounded region, for instance the double cone$^{\, \ref{dc}}$ ${\cal O}_r$,  
and impose in addition a
restriction of the total energy (and thereby also of the total linear
momentum), we get a part of $\cal H$ which we can attribute 
to a bounded region of phase space. 
The restriction of energy can be done, in a somewhat brutal fashion, 
by applying the projection operator $P_E$ for energies below $E$ 
to the set of vectors ${\cal R}({\cal O}_r)_1 \vac$. 
A smooth cutoff function of the energy, such as 
$e^{-\beta H}$ for some sufficiently large positive $\beta$, 
is mathematically more convenient and leads to the subset of 
vectors in $\cal H$
\begin{equation}
{\cal N}_{\beta,r} = e^{-\beta H}\, {\cal R} ({\cal O}_r)_1 \,  \vac .
\end{equation}
It was argued in \cite{HaSw} that a necessary condition for a 
physically reasonable theory 
is the ``compactness criterion''\\[2mm]
$C)$ \ The set of vectors ${\cal N}_{\beta,r}$ is compact in the norm topology
of Hilbert space.\\[2mm]
In other words: for any choice of a positive number $\varepsilon$, the
vectors in ${\cal N}_{\beta,r}$ with norm larger than $\varepsilon$ are
contained in the unit ball of some finite dimensional subspace of
$\cal H$. As $\varepsilon \to 0$, the dimension $N_\varepsilon \to \infty$.

Twenty years later Buchholz and Wichmann \cite{BuWi} 
realized within a different context that the
estimates in \cite{HaSw} could be considerably improved and that the
criterion $C$ should be replaced by\\[2mm]
$N)$ The set ${\cal N}_{\beta,r}$ 
is a nuclear set\footnote{It is
contained in the image of the unit ball of ${\cal H}$ under the
mapping by a positive trace class operator. The trace of this
operator, the ``nuclearity index'', 
is a measure for the size of this set.} for sufficiently 
large $\beta$.\\[2mm]
They argued that this requirement together with
certain bounds on the nuclearity index in its dependence on $r$
and $\beta$ is necessary to ensure known thermodynamic
properties, cf.\ also \cite{BuJu,BrBu} for  
further applications of this condition in the analysis of 
thermal states. 

Several variants of the compactness and nuclearity criteria have been
proposed. We shall not touch here the extensive work about their
relation and consequences. References may be found in
\cite{Ha}. Rather, we shall focus in the following 
on an aspect which emerges from the
foregoing discussion. Irrespective of whether we consider
thermodynamics or particle physics, the von Neumann algebras of all bounded,
contractible regions (such as double cones) are isomorphic. \\[2mm]
{\noindent \subsect The universal structure of local algebras:} \\[2mm]
In \cite{Fr} Fredenhagen 
studied the following geometric constellation: the wedge $W$,
defined in (5.1), and enclosed in it a sequence of double cones  ${\cal
  O}_{r_n}$, tangent to the wedge at the origin, with decreasing radius
$r_{n+1} = \lambda r_n \quad \mbox{for fixed} \ 
\lambda < 1$, so that they contract to the
origin as $n \to \infty$. He found that a non--trivial ``scaling limit''
of the corresponding algebras could only exist if all the 
double cone algebras are of type III$_1$$^{\, {\ref{math}}}$. 
In \cite{BuDaFr} it was shown that the phase space properties 
imply that the local von Neumann algebras are hyperfinite.
Moreover, according to our present 
knowledge, their center is trivial\footnote{The only 
physical reason for the appearance 
of a non--trivial center would be the possibility that superselection rules 
arising from the charge structure might be recognizable already
within a bounded region. But there are good arguments against 
this. Still, a more careful consideration of this in the regime of
local gauge theories might be warranted.}, 
they are ``factors''$^{\, {\ref{math}}}$. 
In \cite{Haa} Haagerup had shown that all hyperfinite factors
of type III$_1$ are isomorphic. Thus we conclude that all local 
algebras are isomorphic to a uniquely defined and well--studied
mathematical object. This emphasizes once more that the physical 
information is not carried by a single algebra. We may compare 
this with the situation in non--relativistic quantum mechanics, 
where we encounter only type I algebras, irrespective of the 
system considered. \newpage
{\noindent \subsect Split property:} \\[2mm] 
Of special interest are inclusion relations between algebras, see
Section 9. We shall address here the case ${\cal R} ({\cal
  O}_1) \subset {\cal R} ({\cal O}_2)$, where the closure of the
region ${\cal O}_1$ is contained in the interior of the  
bounded region ${\cal O}_2$. The  ``split property'' asserts
that then there exist ``intermediate'' factors of type I
such that$^{\ref{math}}$ 
\begin{equation} 
{\cal R} ({\cal O}_1) \subset {\cal N} \subset {\cal R} ({\cal
  O}_2),
\end{equation}
where ${\cal N}$ denotes such a factor (to which we can, however,
not assign a definite localization region). One of the consequences
is the ``statistical independence'' in the situation where 
two regions ${\cal O}_A$ and ${\cal O}_B$ are space--like separated
so that  there is a region ${\cal O}$ which properly contains 
${\cal O}_A$ and is disjoint from ${\cal O}_B$. In this case the 
von Neumann algebra generated by the two local algebras is 
isomorphic to their tensor product. In symbols
\begin{equation} 
{\cal R} ({\cal O}_A) \bigvee {\cal R} ({\cal O}_B)
\simeq {\cal R} ({\cal O}_A) \otimes {\cal R} ({\cal O}_B). 
\end{equation}
This may be regarded as a strengthened form of the locality 
postulate. It tells us that there are states which have no
correlations between the two regions and that the Hilbert space in which 
the two algebras act can be factored into a tensor product 
${\cal H}_A \otimes {\cal H}_B$, analogous to the notion of
subsystems in non relativistic quantum mechanics. This, incidentally,
implies that the discussion of entanglement and non--locality
of EPR--correlations can be done in the same way as in quantum 
mechanics. But the distinction between causal effects, which
are restricted by $L_O$, and EPR--correlations, which 
may persist over (large) spatial distances, is seen more 
clearly in the relativistic setting \cite{SuWe}.

If two observers, nowadays called Alice and Bob, operate in two 
laboratories, there is nothing that Bob can do which 
changes the statistics of any experiment which Alice can make, as long
as they operate at space--like separation. These statistics are 
governed by a well defined 
``partial state'', referring to the lab of Alice, 
which necessarily is impure because it ignores the situation outside
her lab and it does not depend on Bob's activities. 
However, if they look at the statistics of a coincidence
experiment in a state representing a (common) ensemble, there will
in general be correlations in the {\em joint} probability distribution. 

Such correlations persist over large distances if they are 
related to some conservation law. This is not the surprising aspect of 
the EPR--correlations. It is encountered in ``classical''
situations, for instance a state of charge zero decomposing into 
two subsystems $A$ and $B$, carrying opposite charge. Obviously,
if Alice finds positive charge, she knows that Bob must observe
negative charge in a coincidence experiment. The quantum aspect
comes from the possibility that Alice as well as Bob each can decide
to choose among a set of incompatible measurements (which mutually
are compatible). The correlations observed in the pairs of  
such choices are such that they cannot be explained if one tries
to describe the total state as a probabilistic distribution over
pairs of states of ``subsystems'' in any realistic sense. The 
separation of the observation labs does not correspond to a partition
of the system into realistic subsystems. The notion of ``state''
should not be interpreted as the ``mode of existence'' of some
``object'' with ontological significance. It describes 
probability assignments for the occurrence of events (here the clicks 
of detectors in the labs of Alice and Bob). These are real and 
localized.  

The split property has been considered for some time as an additional
assumption and ``standard split inclusions'' have been studied in 
detail in \cite{DoLo}. The recognition in \cite{BuWi,BuDaFr} that it is
a consequence of the phase space properties came as 
a gratifying surprise. If the level density in a theory increases 
too fast, then a finite distance between the boundaries  of the 
regions ${\cal O}_1$ and ${\cal O}_2$ may be necessary for 
relation (5.5) to hold. If the regions have a common boundary 
point, then (5.5)  does not hold.  
%%%%%%%%%%%%%%%%%%%%%%%%%%%%%%%%%%%%%%%%%%%%%%%%%%%%%%%%%%%%%%%%%%%%%%
\section{Charges and statistics}
\setcounter{equation}{0}
The transition from operator algebras to abstract algebras in 
\cite{HaKa} was motivated by the desire for a natural understanding 
of the role of different coherent sectors whose existence had been 
pointed out in \cite{WiWiWi}. Related to this there was the 
challenge to understand the reasons for the Bose--Fermi
alternative. The arguments in quantum mechanics were not adequate
because their starting point, namely the description of a state of
several indistinguishable particles by a wave function in
configuration space, took for granted part of what had to be 
explained. Indeed, other possibilities called
``parastatistics'' had been suggested by H.S.\ Green \cite{Gr}.

In \cite{HaKa} it was argued that the choice of a particular
representation was largely a matter of convenience and that
superselection rules concerning charge and spin must result from 
idealizing the situation at space--like
infinity. Since charges are of eminent physical importance,
the necessary idealizations had to be understood
clearly and related to the appearance of charge sectors and
statistics. The extensive work devoted to this task in the seventies
and eighties need not be reported here in detail. This is described in
Chapter IV of \cite{Ha} and in \cite{BaWo,Ka}, where the pertinent
references may be found.

We shall take here, however, a closer look at the idealizations used
and the ensuing differences in the properties of charges arising from
them. \newpage 
{\noindent \subsect Sharply localizable charges:}\\[2mm]
The approach by Doplicher, Haag, Roberts (DHR picture, \cite{DoHaRo})
started from the idealization that we restrict attention to all those
states $\omega$ which become 
indistinguishable from the vacuum $\omega_0$ by any
observation in the causal complement of a sufficiently large double
cone$^{\, \ref{dc}}$ ${\cal O}_r$. In symbols,
\begin{equation}
\parallel  \omega - \omega_0  
\parallel_{\fA ({\cal O}_r^\prime)} \ \to \ 0 \quad
\mbox{\rm as} \quad r \to \infty, 
\end{equation}  
where ${\cal O}_r^\prime$ denotes the causal complement of ${\cal O}_r$.

It was clear from the outset that thereby one excludes electric charges
from consideration because the flux of the field strength through a
sphere of arbitrarily large radius measures the charge. But it was
believed then that this feature was intimately related to the zero
mass of the photon and could not happen in a theory with a mass gap.
Though it is true that Gauss' law excludes the possibility of a mass
gap \cite{Sw}, 
it turned out that condition (6.1) is too stringent a restriction
even for a purely massive theory (see below).

{}From (6.1) it followed that there are also states in which the
charge is localized sharply in a bounded region. 
This is surprising in view of the discussion in the last section
where we saw that the notion of ``localization of a state'' is in 
general only a qualitative one. Indeed, this feature, following
from the idealization (6.1), will have to be modified, most 
significantly in Section 7. It follows further in this picture that 
the charge structure
arises from the existence of localized automorphisms or endomorphisms
of the net, i.e.\ mappings which preserve the algebraic relations and 
act trivially on the algebras in the causal
complement of some region ${\cal O}$. We call such mappings local 
morphisms. The main consequences of this picture, derived in
\cite{DoHaRo,DoRo}, can be quickly summarized: 
\begin{itemize} 
\item[a)] There is charge conjugation symmetry.
\item[b)] There is a composition law of charges with permutation 
symmetry. It leads to an elaborate calculus of intertwiners between
morphisms localized in different regions.
\item[c)] Each charge is either of bosonic or fermionic type and this
  distinction is reflected in the intertwiner calculus. Note that
this appears as a consequence of the intrinsic structure, outlined
in Section 4, without ad hoc introduction of anticommuting elements.
\item[d)] There is a compact group associated with the charge
  structure. It may be interpreted as the global gauge group.
This is a deep result, obtained by Doplicher and Roberts \cite{DoRo}.
By establishing a novel duality theorem in group theory, they showed 
that the structure of local morphisms and intertwiners, alluded to 
above, defines precisely the dual object of a group. 
\item[e)]Elementary charges correspond to irreducible representations
  of this gauge group. The dimension of the representation corresponds
  to the order of the parastatistics associated with the charge.
\end{itemize}
The simplest case, where the gauge group is the Abelian group $U(1)$, 
gives the well--known situation where all integer values of the charge
appear and we have standard Bose or Fermi statistics. For non--Abelian
groups we get parastatistics. 

These results are put in proper perspective 
if one carries through the same analysis for the 
case  of two--dimensional space--time. There one essential step,
namely the exchange of the position of two space--like separated 
charges by continuous motion, keeping them always space--like, is 
no longer possible and this leads to a much more complicated 
structure \cite{FrReSch,Fro}.
The permutation group is replaced by the
braid group, the Bose--Fermi alternative is changed by the
appearance of anyons and plectons. In the case of less sharply
localizable charges, discussed below, these features appear already
in three--dimensional space--time. By now, there is an extensive 
literature covering these aspects. References and surveys are given in
\cite{Ka,Sch1}. \\[2mm]
{\noindent \subsect Charges localized in space--like cones:} \\[2mm]
Borchers proposed a different selection criterion: Consider all
representations satisfying the spectrum condition $S$. This
also serves to exclude states where the matter density does not vanish at
infinity, but it is weaker than the DHR--criterion. Buchholz and
Fredenhagen used this criterion, restricting the analysis to massive
theories \cite{BuFr}. They found that 
the charges were not necessarily localizable in bounded
regions, there can occur representations where the optimal
localization of charge needs a cone extending to space--like
infinity. 

The results (a) to (e) remain valid in this 
situation \cite{BuFr,DoRo}, but one needs (at least) four--dimensional
space--time to rule out the braid group. Even then, time
honored arguments had to be reexamined if particles carrying such
``BF--charges'' were involved. But it turned out that no serious
changes resulted and even the dispersion relations for S--matrix
elements were not affected \cite{BrEp}. In a massive theory, 
the placement and direction of
the charge carrying cone plays no role for the unitary equivalence 
class of the representation (the superselection sector).
Only the topological property that the sphere at space--like
infinity has to be punctured somewhere is relevant.\newpage 
{\noindent \subsect Absence of mass gap:}\\[2mm] If there 
are excitations of arbitrarily small energy, then the
description of the set of  superselection rules becomes a formidable
task due to ``infrared clouds''. In the case of QED the discussion of
this \cite{Bu2} led to some interesting consequences. The optimal
localizability of charge in a space--like cone remains, but different
placements of this cone correspond to different infrared clouds and
thereby to unitarily inequivalent representations. 

Since a charged particle is always accompanied by an infrared cloud
which depends on its state of motion, its mass is not sharply 
defined and represents only a lower bound of the energy--momentum spectrum 
(infraparticle problem \cite{Sch,FrMoSt,Bu3}). But it is possible to
give a precise meaning to the notion of 
``improper state of a charged particle
with sharp four--momentum'' (a generalized Dirac ket) as a ``weight'' on the
algebra $\fA$. In contrast to the case of neutral particles, 
a superposition of these improper states to form 
a wave packet with specified localization properties
is not possible. Nevertheless, a clear formulation
of collision theory for charged particles and hard photons is
available \cite{BuPoSt}. \\[2mm]
{\noindent \subsect Summary and questions:} \\[2mm]
The analysis described in this section started from the aim of
understanding the superselection structure. In the massive case it
led to the appearance of charge quantum numbers related to
equivalence classes of local morphisms (charge creation). The
laws of composition and conjugation of these morphisms miraculously
turned out to correspond precisely to the dual object of a compact
group (global gauge group). Locality leads to a permutation symmetry
whose implementation demands (para) Bose or Fermi
statistics. If, instead of Minkowski space, one considers theories in
lower dimensions, then these statements must be modified.

But there is evidence that the charge structure has deeper roots and is
not necessarily reflected by superselection rules. In the Standard
Model it is associated with a ``principle of local gauge
invariance'' which has not been incorporated in the scheme discussed
so far. Superselection rules and observable charges appear only as
the survivors in the ``unbroken part'' of a very large symmetry. 
The notion of ``spontaneous symmetry breaking'' relates to
the possible existence of different ``phases'' with different behavior
at space--like infinity, e.g.\ ``long range order''.
The observable charges may depend on the phase. We shall address the
meaning of ``symmetry'' in general and of the ``local gauge
principle'' in particular in the next section.
%%%%%%%%%%%%%%%%%%%%%%%%%%%%%%%%%%%%%%%%%%%%%%%%%%%%%%%%%%%%%%%%%%%%%%%%%%%%%
\section{Symmetries, local gauge principle}
\setcounter{equation}{0}
The word ``symmetry'' has several connotations. In $A_3$ we used
it in the ``active'' sense. An element of the symmetry group changes
the physical situation, noticeable by an observer, to an equivalent 
situation. By ``change of situation'' we mean in standard quantum
theory that the state and all observables are
altered. ``Equivalent''  means that all laws of nature apply in
unchanged form in the new situation. More often, however, symmetry is
understood in the ``passive'' sense as providing alternative
descriptions for the same physical situation, expressing the fact that
the (known) laws of nature do not distinguish a preferred way of
coordinatization within a class of equivalent ones. In either case, we
have to consider reference frames (coordinate systems) 
for the symmetry group and for the
objects on which it acts. 

In the case of the Poincar\'e symmetries we have implicitly assumed
that an observer can establish a global reference frame in Minkowski
space (fixing a point as the origin and a Lorentzian tetrad). He
does this with the help of some macroscopic bodies and clocks which
are not included in the physical system considered in the
theory. (They are part of the ``observer side'' of the
Bohr--Heisenberg cut.) Then, keeping this reference frame fixed, he
can interpret the Poincar\'e transformations in the active sense. We
shall accept this idealization here and ignore its limitations
(indicated by general relativity on the one hand and the quantum
nature of bodies used in the establishment of the frame on the other
hand). 

The active interpretation is, however, not possible 
in many cases where the symmetry we speak about is more indirectly
inferred (or assumed) and the macroscopic objects available do not define
a reference frame which the observer could control and
change. Thus the global gauge groups mentioned in the last section may
be regarded as describing a symmetry of the theory. But the objects
on which the group acts cannot be measurable quantities in the sense  
of standard quantum theory. They cannot be accommodated in the algebra
of observables because there is no operational way of establishing
a reference frame relating to this group and thus the observables 
cannot depend on it. They must be invariants. One can extend the 
net of observable algebras $\RO$ to a net of ``field algebras'' $\FO$
on which the global (compact) gauge group $G$ acts \cite{DoRo}. 
But linear relations between elements of $\FO$ which transform 
according to inequivalent representations of $G$ are void of 
physical meaning. This reflects the limitations of the superposition 
principle \cite{WiWiWi} and the related feature that the causality 
requirement, expressed by $L_O$, does not apply to the net $\FO$.

Apart from the possibility of fixing a reference frame in an 
operational way, there is another problem whose recognition
was one of the 
keys leading to the development of the theory of general relativity:
The comparison of reference systems used by observers in different
regions is ambiguous because it requires some bridge connecting them 
(some transport of information \cite{Ma}) and
the choice of the bridge plays a role. If we regard {\em local}
gauge transformations as an internal symmetry, then both aspects enter. There
is no observable way to fix a reference frame and there is no
unambiguous way of comparing frames in different locations. 
The observables must be independent of the choice of these frames.
Thus, to incorporate the local gauge principle by specifying an
internal symmetry group, we must again augment the algebraic scheme.

A symmetry is expressed by a  mapping of the mathematical structure 
onto itself. In the case of the structure
outlined in \secfouandfiv, this means that it is 
described by an automorphism $\alpha$ of the algebra $\fA$
which conserves the net structure. In other words, the image
of the algebra $\RO$ must again be the algebra of some 
space--time region ${\cal O}_\alpha$,
\begin{equation}
\alpha \RO = {\cal R} ({\cal O}_\alpha).
\end{equation}
Since this should hold for arbitrarily small
regions, the map 
${\cal O} \rightarrow {\cal O}_\alpha$ must result from a point
transformation $g_\alpha$,
\begin{equation}
x \rightarrow g_\alpha \, x,
\end{equation}
which furthermore has to conserve the causal 
structure in Minkowski space. For a precise discussion see \cite{Ar}. 
This limits  $g_\alpha$ to the elements of the Poincar\'e group, possibly
extended by dilations. Apart from these 
``geometrical symmetries'', which change the regions, there may be also
internal symmetries, corresponding to automorphisms transforming each 
$\RO$ onto itself,
\begin{equation}
\alpha \RO = {\cal R} ({\cal O}),
\end{equation}
a prominent example being charge conjugation.  

An automorphism of $\fA$ is called ``inner'' if it is implemented
by a unitary element $U$ belonging to the global C$^*$--algebra 
$\fA$, 
\begin{equation}
\alpha A = U A U^{-1}, \ A \in \fA.
\end{equation}
None of the global symmetries can be inner since they act
non--trivially
on observables localized arbitrarily far, whereas $\fA$ contains 
only quasilocal elements. However, it appears that a much more 
important notion is local implementability of (possibly only local)
symmetries. This means that we focus attention on the action 
of $\alpha$ on some chosen algebra $\RO$. Then $\alpha$ may be
called ``locally inner'' if there exists some finitely extended region
$\widehat{\cal O}$ such that 
\begin{equation} 
\alpha A = U A U^{-1} \quad  
\mbox{for some} \ U \in {\cal R} (\widehat{\cal O}) 
\ \mbox{and all} \ R \in {\cal A} ({\cal O}). 
\label{implementation}
\end{equation} 
It is then important to characterize a lower bound for the choice of 
$\widehat{\cal O}$. If the symmetry shifts ${\cal O}$ to ${\cal
  O}_\alpha$,
then the split property implies that one may choose for $\widehat{\cal O}$ in 
(\ref{implementation}) any connected region which contains the closure of
the causal completion of ${\cal O} \cup {\cal O}_\alpha$. 

This entails the following 
analogue to Noether's theorem in the algebraic setting:  
The infinitesimal generator of any continuous symmetry 
is locally implemented by a Hermitian operator which is affiliated with a
slightly larger region \cite{BuDoLo}. 
(If the charge structure is adequately described by the considerations
in \secsix, then this holds likewise for 
the global gauge symmetries acting on the field algebra.) 
One may expect that these local generators 
determine a pointlike field (density) $\rho$ in the
limit as ${\cal O}$ shrinks to a point (cf.\ \secfou).
Moreover, if the symmetry commutes with time translations, there should
hold a continuity equation,  
\begin{equation}
{\dot{\rho}} = \mbox{div} \, \bfj,
\end{equation}
where $\bfj$ is again a Wightman field. As discussed in 
\cite{BuDoLo}, there remain some unresolved ambiguities 
in carrying through this intuitive argument whose
significance is not yet properly understood. But 
these remarks may indicate the role of pointlike fields, such as 
the components of the energy--momentum tensor and certain currents
within the general scheme.

Let us finally discuss the case of internal symmetries relating  
to the local gauge principle. There the mathematical structure 
referring to the non--invariant elements is much  
more subtle. To fix ideas, let us think of an internal symmetry
group like SU(2) or U(1). In the classical theory, the appropriate
structure is described by fiber bundles in which the notion of a field
configuration in some region ${\cal O}$ is replaced by that of a 
section in some (associated) bundle. The section obtains
physical relevance only in conjunction with a connection. This
demands that we endow the region ${\cal O}$ (assumed to
be contractible) with a collection $C$ of paths linking a fixed 
reference point $x_0 \in {\cal O}$ uniquely 
to every other point $x \in {\cal O}$.
Secondly, that we attach a ``charge transporter''  
$\Gamma_{{x}^{}_{0} \hspace*{0.7pt} x}$ 
to each such path with the help of the connection form.
Using this device, we can compare the elements in different 
fibers and thus 
obtain the analogue of an ordinary field, say
$\varphi_C (x)$, for which the algebraic operations of addition and
multiplication at different points are
meaningful and for which the transformation by elements of the
symmetry group, referring now to all of ${\cal O}$, is defined.

In adapting this to the non--commutative situation, we meet several 
difficulties. The first concerns the proper assessment of the singular
quantities associated with points and lines. Since the connection forms 
are no longer ordinary functions, there is the question of
whether the  $\varphi_C (x)$ can still be regarded as ``operator--valued
distributions'', as assumed for ordinary fields in \sectwo. This is
presumably not the case. 
But let us for the moment ignore this problem and proceed as in
Section 4. This would lead us to local field algebras 
${\cal F}_C ({\cal  O})$. The fixed points within ${\cal F}_C ({\cal O})$ 
under the action of the group 
could be interpreted as elements of the observable algebras  
${\cal R} ({\cal O})$ which have significance without reference to
the choice of $C$. If furthermore we could find within 
${\cal F}_C ({\cal O}) $ subspaces of partial isometries 
transforming under a specific irreducible representation 
of the symmetry group, then we could define endomorphisms for $\RO$,
in complete analogy to the case of global gauge symmetries.
We would thus arrive in the area 
studied in \cite{DoRo} with one difference: instead of
``localized endomorphisms'' for all of $\fA$, we would have to
consider now endomorphisms restricted to the algebra 
${\cal R}({\cal O})$ for
some region. Also, we would follow the path in \cite{DoRo} in the
opposite direction. Instead of starting with the endomorphisms and
their intertwiners and ending with a group, we would start from the
group from which the dual structure emanates.

The purpose of the excursion in the last paragraph was just to
indicate some parallelism of the superstructure met in quantized gauge
theories with that found in theories with a global gauge group. 
The main problem with this
``as if'' picture comes from the feature that we cannot remedy the
singular nature of $\varphi_C (x)$ just by smearing out with a test
function $f(x)$, keeping $C$ fixed. On the other hand, the algebraic
relations between objects referring to different choices of $C$ are
not meaningful. Work in lattice gauge theory \cite{Sei} and perturbation 
theory \cite{Ste} indicates, however, that ``quantum charge 
transporters'' $\Gamma_{x  y}$ may be definable as distributions 
in $x$ and $y$. These objects, corresponding in the field theoretic
setting to finite Mandelstam strings, would allow us to construct by 
algebraic operations special elements in the observable algebras $\RO$
for which we can distinguish two kinds of supports: ``charge supports'',
relating to the supports of the test functions in $x$, respectively
$y$, used for the smearing of 
$\Gamma_{x  y}$, and a 
``causal support'', involving in addition a bridge region between
the charge supports. The intrinsic significance of the notion of ``connection''
should then be understood by studying the effect of cutting such
objects between disjoint charge supports.

Let us add one comment. In general the group structure just
distinguishes conjugacy classes within the group so that a reference
frame is also needed to characterize individual group elements. In the
case of an Abelian group this is not necessary. A conjugacy class
consists only of a single element. This brings some
simplifications since the symmetry may be locally implementable
by an observable field $\rho$ (the charge density
in QED). The trivial action of the group on the observables is then
expressed by Gauss' law, i.e.\ the existence of a local observable
field ${\bf E}$ such that 
\begin{equation}
\rho = \mbox{div} \, {\bf E}.
\end{equation}
As already mentioned, the consequences of this feature have been
studied extensively.

There are many aspects of the local gauge principle we cannot touch here 
(mostly because they are not worked out in adequately clear form). The
elaboration of this rich structure to a degree of conciseness
comparable to that of the previous sections appears to us as a task
worthy of the sweat of the noble. It is presumably an essential step
towards the characterization of a specific theory within the general
frame described in previous sections along the lines suggested by the
progress of high energy physics in the past decades. 

%%%%%%%%%%%%%%%%%%%%%%%%%%%%%%%%%%%%%%%%%%%%%%%%%%%%%%%%%%%%%%%%%%%%%%%%%%%%
%%%%%%%%%%%%%%%%%%%%%%%%%%%%%%%%%%%%%%%%%%%%%%%%%%%%%%%%%%%%%%%%%%%%%%
%%%%%%%%%%%%%%%%%%%%%%%%%%%%%%%%%%%%%%%%%%%%%%%%%%%%%%%%%%%%%%%%%%%%%%
\section{Short distance structure}
\setcounter{equation}{0}
The previous sections were concerned with the development of a
conceptual frame and corresponding mathematical structure, which can
be accepted as reasonably complete and simple and provides natural
answers to a variety of questions coming mainly from quantum field
theory. But it is, of course, of paramount importance to see how a
specific theory can be characterized within this general frame. {}From
the observation that the local gauge principle together with a
postulate of ``minimal coupling'' and some knowledge about the
relevant degrees of freedom fixes the choice of a Lagrangian in
classical field theory almost uniquely, we may surmise that the
information which is needed to define a theory, beyond that supplied
already by general principles and specification of internal
symmetries, is encoded in the short distance structure.\\[2mm]
{\noindent \subsect Scaling algebras:}\\[2mm]
Renormalization group methods have proven to be a powerful tool for
the classification of this structure in quantum field theory. So it is
gratifying that they have a very simple counterpart in the algebraic
approach \cite{BuVe}. The essential idea is to consider functions $\sA$ of
a scaling parameter $\lambda \in \RR^+$ with values in the
algebra of observables. In other words, the ``value'' $\sA(\lambda)$ is
an element of $\fA$. These functions form, under the obvious
pointwise defined algebraic operations, a normed algebra
$\sfA$ on which the Poincar\'e transformations
$(x,\Lambda)$ act by automorphisms 
$\sa_{\, x,\Lambda}$ related to those in $\fA$ by
\begin{equation}
( \sa_x \, \sA )(\lambda) = 
\alpha_{\lambda x} \, \sA(\lambda) , \quad
( \sa_\Lambda \, \sA )(\lambda)= 
\alpha_\Lambda \, \sA (\lambda).
\end{equation}
The norm is defined by
\begin{equation}
\parallel \sA \parallel = \sup_\lambda \parallel \sA(\lambda) \parallel.
\end{equation}
The local structure of the original net is lifted to $\sfA$ by setting
\begin{equation}
\sfA({\cal O}) = \{ \sA : \ \sA(\lambda) \in  
{\cal R} (\lambda {\cal O}), \ \lambda \in \RR^+ \},
\end{equation}
and the momentum space properties of the elements $\sA$ of the scaling algebra
are controlled by the requirement that 
$\sa_x \, \sA $  and $\sa_\Lambda \, \sA$ depend norm--continuously 
on $x$ and $\Lambda$, respectively. The latter condition entails 
that the values $\sA(\lambda)$ of $\sA$, being localized in
$\lambda {\cal O}$, have a momentum transfer 
of order $\lambda^{-1}$, in accord with 
the uncertainty principle. In this way 
one obtains a local, Poincar\'e covariant net
which is canonically associated with the original theory.

One may regard the values $\sA(\lambda)$ as observables in the theory
``at scale $\lambda$'' corresponding to a change of the original unit
of length and thereby of the metric tensor by the factor
$\lambda$. The graph of a function $\sA$ establishes a relation between
observables at different scales, in analogy to renormalization group
transformations. However, in contrast to the field theoretic setting,
there is no need to identify individual observables at different
scales. All functions satisfying the constraints indicated above are
admitted.  It may seem strange at first sight that with such loose
constraints the net $\sfA$ could provide any interesting
information. But this may be understood by recalling that  
the relevant physical information is contained in the net
structure and only the identification of the sets of operators associated to
regions is necessary. For that reason one has much more
freedom in choosing the relation between observables at different
scales. 

The next step is to describe the states in the theory at 
given scale $\lambda$ with the help of the scaling algebra. 
This can be done by lifting the states $\omega$ of the 
underlying theory at scale $\lambda = 1$ to $\sfA$, setting
\begin{equation}
\underline{\omega}_{\, \lambda} (\sA \, \sB \cdots \sC) = 
\omega(A(\lambda)  \, B(\lambda) \cdots C(\lambda)). 
\end{equation}
The short distance 
properties of the theory can then be analyzed by proceeding to 
\begin{equation}
\underline{\omega}_{\, 0} (\sA \, \sB \cdots \sC) = 
\lim \, \omega(A(\lambda)  \, B(\lambda) \cdots C(\lambda)), 
\end{equation}  
where we understand the symbol $\lim$ as denoting any limit point of the
sequence on the right hand side for $\lambda \to 0$. The existence of such
limit points is guaranteed by general mathematical
theorems\footnote{To use a favorite expression of R.V.\ Kadison:
``Highly efficient abstract nonsense''.  
Specifically, it is the weak compactness of the
unit ball in state space which is used.}. Any such limit point is 
a pure vacuum state on $\sfA$, irrespective of the state $\omega$ 
from which one starts.

By the GNS construction one obtains from
$\underline{\omega}_0$ a representation of a local net of von 
Neumann algebras, acting in
a Hilbert space, which we call the scaling limit of the
theory. Three distinct possibilities can arise. The limit may yield a
classical theory (commutative algebras). This arises when all 
functions $\sA$ become multiples of the unit element
as $\lambda \to 0$. Secondly, there may exist
many different limit theories (indicating the presence of 
an ``unstable ultraviolet fixed
point''). The third alternative that the limit points (8.5)
define a unique theory and that this is not
classical is, of course, the
most interesting one. It may be regarded as a distinctive mark
characterizing renormalizable theories 
with a stable ultraviolet fixed point in an intrinsic way,
i.e.\ without reference to perturbation expansions or other
approximation methods. 
One may expect that the scaling limit theory is 
simpler than the original one; in the extreme case it may turn out to
be a theory of free fields (asymptotic freedom) \cite{Bu4}. 

This suggests that it is a reasonably well defined mathematical
problem to investigate in the algebraic setting 
the existence and uniqueness of theories which
have prescribed symmetries in the sense of the preceding section and
are asymptotically free. The aims of such an enterprise would 
be similar to those pursued in ``constructive quantum field 
theory'', but the methods may complement  previous efforts 
\cite{GlJa}. \\[2mm] 
{\subsect Germs of states:} \\[2mm]
There is another way to look at the short distance structure
\cite{HaOj} which
yields a somewhat different type of information, relating to point
fields and operator product expansions \cite{WiZi}. Any state $\omega$
has (as an expectation functional) a restriction to each subalgebra $\RO$,
called the partial state in 
${\cal O}$. For given $\RO$ we may consider the set
of the corresponding partial states, or rather its complex hull
$\SO$, which is a Banach space with distinguished positive
cone. The maps ${\Sigma} ({\cal O}_2) \to {\Sigma} ({\cal O}_1)$,  
which are obtained by restricting the functionals in 
${\Sigma} ({\cal O}_2)$ to ${\cal R}({\cal O}_1)$
if ${\cal O}_1 \subset {\cal O}_2$, induce the  
structure of a presheaf on the collection of $\SO$:
A partial state on ${\cal O}_1$ corresponds to an equivalence
class of partial states in ${\cal O}_2$. One may define
an equivalence relation with respect to a point $x$,
\begin{equation}
\psi \quad
\mbox{$\simeq$ \raisebox{-1.4ex}{\hspace*{-2.9ex} $\scriptstyle x$}}  
%\stackrel{x}{\simeq} 
\quad \psi^\prime,
\end{equation}
meaning that there exists some neighborhood of $x$ in which
the restrictions of $\psi$ and $\psi^\prime$ coincide.
We shall call such an equivalence class $\{ \psi \}_x$ 
a ``germ at the point $x$''.

The nuclearity property,   
discussed in Section 5, suggests that one may obtain a
tractable description of such germs in the following way: Focus
attention on functionals $\psi_E$ with total energy below $E$ and
restrict them to the algebras$^{\, \ref{dc}}$ ${\cal R} ({\cal O}_r)$. 
The resulting spaces are denoted by $\Sigma_E ({\cal O}_r)$.
Disregarding functionals of small norm, each of them  
is finite dimensional. A measure for the accuracy of  
these finite approximations is the distance $\delta_n (E, r)$ 
between the unit balls in $\Sigma_E ({\cal O}_r)$  
and in the closest $n$--dimensional subspace of functionals. 
This distance decreases as $r \to 0$ and increases as 
$E \to \infty$; moreover, for fixed $E,r$ it decreases 
with growing $n$.    

Under reasonable assumptions, the distance functions $\delta_n (E, r)$ 
vanish with increasing $n$ of increasingly high order as
$r \to 0$. In the example of the theory of  
a free scalar field $\varphi_0$ one has for $E r < 1$ 
\begin{equation}
\delta_n(E,r) \simeq (E r)^{d_n}, 
\ \ \mbox{where} \ d_1 = 1, \ d_2 = 2, \ 
d_3 = \dots = d_7 = 3 \   \mbox{etc.} 
\end{equation}
Proceeding to the dual picture (the co--germs, which are  
associated with the algebra), this gives an
increasing number of pointlike fields which are needed to distinguish
the functionals in $\Sigma_E ({\cal O}_r)$  
with increasing accuracy $\delta_n(E,r)$. In the free scalar theory, 
the unit operator corresponds to $n=1$. For $n=2$, the field 
$\varphi_0$ is needed also, for $n \leq 7$
the 4 derivatives $\partial_\mu \, \varphi_0$ enter and the Wick power
$: \! \varphi_0^2 \! :$. 
Ultimately, all elements of the Borchers class appear. The
field equations give a reduction in the number of new independent
elements for increasing $n$. Thus one has an ordering of the elements
of the Borchers class according to their significance in the regime
of small $E r$ and one may consider approximation schemes corresponding
to operator product expansions.

\section{Inclusions}
There are two distinct types of questions in which the study of
inclusions of algebras plays a role. The obvious one in our context
comes from the inclusions of regions in space--time. The less obvious
one from endomorphisms of the algebra $\fA$. We shall begin with
the latter because it provides another surprising example for
``prestabilized harmony'' between physics and mathematics.

The analysis in Section 6 related charge creation to the existence of
endomorphisms of $\fA$. This led to the recognition that,
associated with each type of charge, there is a ``statistics parameter"
$\lambda$ which, in the case of 
four--dimensional space--time, could only
take the values $\pm n^{-1}$, where $n$ is an integer,
the statistics dimension, and the sign
distinguishes the Bose and Fermi case \cite{DoHaRo}. 
In two--dimensional space--time
there is a much wider range of possibilities. The ``statistics
dimension'' $|\lambda|^{-1}$ can take non--integer values and, instead
of a sign, complex phase factors can appear. Instead of the Bose--Fermi
alternative one has ``braid group statistics'' \cite{FrReSch}.

Motivated by a quite different circle of questions in mathematics, 
Vaughan Jones discovered that for certain inclusions of type
II factors there exists an ``index'' which can take only a restricted
set of values and that there is a relation of this structure with
representations of the braid group. This initiated a mathematical
development leading for example 
to substantial generalizations and applications to
the theory of knots. It took several years till the close connection
of these mathematical developments with the composition laws of charge
quantum numbers was recognized by Longo \cite{Lo}, who showed that 
the statistical  dimension is a 
(generalized) Jones index. While in
four--dimensional space--time the latter is restricted to integers and this
feature is connected with the permutation group, the full complexity
appears in the analysis of possible charge structures in two--dimensional
space--time.

Coming now to the inclusions in the net of algebras, we may restrict
ourselves here to a few remarks since this is extensively discussed in
the contribution of Borchers to this issue. The central result is that
a few algebras suffice to determine the whole net as well as the operators
representing the Poincar\'e  group and the TCP--operator. This amazing
fact, recognized by Wiesbrock 
\cite{Wi,KaWi} using basic results of Borchers
\cite{Bo2},  can be made plausible intuitively by starting from the
discovery of Bisognano and Wichmann \cite{BiWi} which indicated that the
modular automorphism group for the vacuum state and the wedge region
(5.1) gives the Lorentz transformations in the $x_0$--$x_1$--plane. If one
takes a second wedge, included in the first one, one obtains
``half--sided modular inclusions'' for which the modular
operators and conjugations generate a whole family of algebras and the
translation operator in a light--like direction in the
$x_0$--$x_1$--plane. Repeating this construction, changing $x_1$ to $x_2$ and
$x_3$, the whole net is obtained by intersections and the full
Poincar\'e group is obtained.

One comment should be added. Each of the six algebras used in the
construction is isomorphic to the unique hyperfinite type
III$_1$--factor. Their association with specific regions in Minkowski
space could be regarded as secondary. It only serves to fix a general
relation between these algebras. So one may replace the assumption
$L_O$, implying that the label 
${\cal O}$ in the basic correspondence (4.1) 
should be interpreted as a region in Minkowski space, by a weaker
one \cite{BuDrFlSu}. 
The main structural relations needed refer to inclusions and
complements and these are directly encoded in the algebraic relations.

%%%%%%%%%%%%%%%%%%%%%%%%%%%%%%%%%%%%%%%%%%%%%%%%%%%%%%%%%%%%%%%%%%%%%%
%%%%%%%%%%%%%%%%%%%%%%%%%%%%%%%%%%%%%%%%%%%%%%%%%%%%%%%%%%%%%%%%%%%%%%
\section{Summary, comments, conclusions, perspectives}
\setcounter{equation}{0}

Looking back at the understanding of relativistic quantum 
physics fifty years ago, it may be fair to say that the second
half of the century brought no revolution comparable in impact
to the radical changes of basic concepts which shook the first
three decades of this century. It was a period of steady evolution, 
but it led to significant changes of perspective.

We discussed here the synthesis of quantum theory and special 
relativity, incorporating some information from other sources 
and striving to bring out the essentials of a coherent conceptual
and mathematical structure. On the side of quantum theory, we 
started from the orthodox position, distinguishing the observer
and his instruments from the ``physical system'' which here, in 
principle, could be the whole universe minus the observer. We also
adopted the point of view of Heisenberg and Dirac that observables 
and manipulations of the system by the observer are mathematically
described by elements of a non commutative algebra and that the 
(abstract) algebraic relations constitute the essence of the theory.
Keeping in mind Niels Bohr's message that we must be able to 
``tell our friends what we have done and what we have learned''
and his conclusion that this forces us to describe the side of the
observer in the ``language of classical physics'', we note that indeed
we retain one classical anchor, namely classical space--time in 
which we describe the placement of instruments and the Poincar\'e 
symmetry, used in the active sense of pushing around instruments. 

The bridge between the mathematical formalism and the communication
with our friends is provided by the correspondence (4.1) between
space--time regions and algebras and by the realization of the 
Poincar\'e group by automorphisms (7.1), (7.2) of the net of local algebras.
Its combination allows us to describe geometric constellations of
instruments and to analyze the energy transfer between the 
instruments and the ``system'', which suffices for a full
physical interpretation of the information contained in the
mathematical scheme.  

How much do we know about the net of algebras? The two central 
pillars come from the relativistic causal structure of space--time,
expressed by the postulate $L_O$ of \secthr, and from the 
stability requirement, expressed as postulate $S$ in \sectwo.
But the closer study in \secfiv\ leads to important refinements. 
On the one hand it shows that if we focus attention only on regions
of finite extension, there is a faithful representation of the abstract
algebraic elements by operators acting on a Hilbert space $\cal H$.
Each individual subalgebra $\RO$ is 
isomorphic to a universal, 
known and well--studied object: the unique hyperfinite factor of 
type III$_1$. 

On the other hand, the consideration of different classes of states 
brings out clearly that the abstract algebras are the basic objects
whereas their representations in terms of operators in Hilbert space
is a matter of convenience which may be adapted to the situation 
under consideration. Inequivalent representations of one and the 
same net of abstract algebras describe different idealizations which
are useful in different regimes. Prime examples are thermal  
equilibrium states and states carrying some (global) charge quantum
number (\secsix). 

Postulate $S$ is strengthened to the nuclearity postulate $N$, 
implying roughly that finite phase space volumes correspond to finite
dimensional subspaces of $\cal H$. The locality principle $L_O$
is strengthened to the ``split property'' (5.5) which allows a
factorization of $\cal H$ for disjoint, space--like separated 
regions in analogy to the notion of subsystems in non relativistic
quantum mechanics. This, incidentally, implies that the discussion
of entanglement and non--locality of the EPR--correlations can be 
done in the same way as in quantum mechanics. 
But in this setting the distinction between causal effects whose
propagation is limited by light cones, as demanded by  $L_O$, 
and EPR--correlations, which may persist over (large) spatial
distances,  is seen clearly. \\[1mm]
{\it Comment:} If instead of Minkowski space one considers 
curved space--time, then the algebraic part of the theory carries over
smoothly since the net structure refers only to inclusion relations
and causal complementation and both remain well defined if the metric
structure is classically given in terms of a gravitational background
field \cite{Di}. The loss of Poincar\'e symmetry demands, however, 
that the stability requirement $S$ must be replaced. Suggestions of 
how this may be done have been proposed. See e.g.\
\cite{Wa,BrFrKo,Ha,BuDrFlSu}. 
The most interesting physical consequences which can be treated 
in this setting are the 
Bekenstein temperature and Hawking radiation associated with black 
holes. One should note, 
however, that our present understanding of the stability requirement 
is not fully satisfactory.

We must now face up to the essential task of defining one specific
theory within the still rather general frame. The most significant
progress in high energy theory in the past decades has been the 
development of the Standard Model. It combines the choice of specific
internal symmetry groups with the sharpened locality principle which,
more than eighty years ago, had led to the general theory of
relativity: there is no preferred global reference frame; the relation
between frames in different locations depends on the choice of 
a path connecting them.

But here and now we talk about reference frames for the degrees of 
freedom associated with internal symmetries, not about the frame for 
space--time coordinates. The incorporation of internal symmetries 
subject to this ``local gauge principle'', which demands that there is
no preferred global reference system for them, 
is addressed in \secsev. It is not a straightforward task to
transfer the notions of sections and connections, familiar from the 
classical formulation with fiber bundles, to the quantum level.
We briefly sketched an approach which could lead to an intrinsic
understanding of the meaning of quantum connection. One essential
aspect appears to be that in addition to the causal support ${\cal  O}$, 
used in the correspondence (4.1), one must introduce finer
distinctions in ${\cal R} ({\cal O})$ by so--called ``charge 
supports''. Specifically, one needs special elements in 
${\cal R} ({\cal O})$
with disjoint, complementary charge supports related to representations
of the gauge group. 

Much remains to be done in the development of such
ideas till a concise and complete structure is reached. But it is an
effort well worth--while since it can open the gate to a very wide
field. We mentioned already the need for a good definition of a
specific theory along the lines suggested by the Standard Model.
Combined with the short distance analysis 
(\seceig), the question of existence and uniqueness could be
approached in precise mathematical terms. But beyond that let us 
mention some old dreams: the supersymmetric unification of internal
and geometric symmetries, treated as local gauge symmetries. 
This may, in fact, even suggest a natural approach to the synthesis of
general relativity and quantum physics since, once we sacrifice 
the global nature of translations, we may also treat the 
Lorentz group as an internal SL(2,C)--symmetry. 

None of these perspectives is of a truly revolutionary nature. They
constitute a natural development of existing ideas. But it seems that 
this development has not yet reached its end, its essential limits.
The road from QED to QCD exemplifies that old equations and
principles, properly understood and adapted, contain a lot of relevant
new physics. The problems mentioned above indicate the wide range 
of efforts still needed to clarify and round off this era. 

At the same time one cannot ignore the signs indicating the approach
of some radical change in basic concepts. What will be the role of 
space--time in the paradigm of a future theory and how will the
orthodox position of quantum theory be affected? It is already evident
that the classical anchor, provided by the operational interpretation
of space--time as the bridge between the mathematical formalism of the
theory and the simple language needed for ``telling what we have 
learned'' cannot be pushed to extremes. We do not place and control
instruments in regions of $10^{-16}$ cm extension. If we look for
a synthesis of quantum physics with general relativity, for instance
along the lines indicated above, then this means that we introduce on
the side of the mathematical structure of the theory substantially
more detailed ontological extrapolations than can be directly related
to observations. The needed bridge on which Bohr insisted (for good
reasons) must be established on an intermediate level, such as the 
definition of some (classical) background which must first be derived 
from the theory as an approximation under suitable circumstances. 

There arises the question of how we can (within the scope of physics)
divide the universe into distinct, individual parts to which we can
give a name. This is indeed the main message brought home by the
EPR--type experiments, because we see that the notions of ``system''
and ``state'' are approximate or relative unless we consider the whole
universe as the system. As long as we regard space--time as a pregiven
continuum, we may use this for the purpose of subdivision. 
If we give up this anchor, then what remains?

If we believe in a fundamental indeterminism of the
theory, then we must distinguish between the realm of facts and the
realm of possibilities, represented by probability assignments. The 
former is, at present, reduced to ``observation results'', the latter
to the notion of ``state''. Strictly speaking, an observation result is 
a macroscopic change which enters the consciousness of several human
beings. This is certainly necessary for testing a theory. But hardly
as a basic concept. Can it be generalized by the notion of an 
``event'' which does not depend on the senses and consciousness of 
humans? Is the role of space--time ultimately just 
the set of relations within a pattern of events? Does the distinction between 
potentialities and facts imply a fundamental significance of the arrow
of time? Facts belong to the past, possibilities to the future. 
For some tentative steps in such directions compare \cite{Ha2}.

Let us conclude this essay with the acknowledgment that there 
remain many questions and we are very far from a ``theory of
everything''.\\[3mm]
{\noindent \large \bf Appendix} \\[3mm]
\setcounter{section}{1}
\setcounter{equation}{0}
\renewcommand{\theequation}{\Alph{section}.\arabic{equation}}
For the convenience of the reader, we collect here some 
facts and notions from the theory of operator algebras which 
are used in  the main text.

In the algebraic approach to quantum theory, the basic 
mathematical objects are C$^*$--algebras. A C$^*$--algebra
$\fA$ is a complex linear space, equipped with an 
associative product, a 
$*$--operation (defining the adjoint) and a distinguished 
norm $\parallel \cdot \parallel$. With respect to the 
corresponding norm topology, $\fA$ is complete, i.e.\ a Banach space. 
We assume that $\fA$ contains a unit
element $1$.

A state $\omega$ on $\fA$ 
is a complex linear functional which 
attains non--negative values 
on all elements of the positive cone 
$\fA_+ = \{ A^* A : A \in \fA \} \subset \fA$ 
and which is normalized, $\omega (1) = 1$. 
It is a basic fact, established by Gelfand, Naimark and 
Segal (GNS construction), that any state $\omega$ determines
(a) some Hilbert space ${\cal H}$, (b) a mapping 
$\bpi$ from $\fA$ into the algebra of bounded linear 
operators on ${\cal H}$ which preserves the algebraic 
relations (i.e.\ a homomorphism) and (c) some normalized vector 
$\Phi \in {\cal H}$ such that 
\begin{equation}
\omega (A) = (\Phi, \, \bpi (A) \, \Phi)
\quad \mbox{for} \quad A \in \fA.       
\end{equation}
In this way, $\fA$ is mapped to a concrete
C$^*$--algebra $\bpi (\fA)$ of Hilbert space operators and 
the state $\omega$ is interpreted as an expectation functional 
on $\bpi (\fA)$. 

On a Hilbert space one can introduce the 
notion of weak convergence of sequences of bounded operators 
(all matrix elements of the sequence converge). The 
resulting limits are again bounded linear operators.  
It is therefore meaningful to proceed from 
$\bpi (\fA)$ to its ``weak closure'' 
${\cal R} = {\bpi (\fA) }^-$, i.e.\ the set of operators 
consisting of  $\bpi (\fA)$ and all weak limit points. 
The set ${\cal R}$ is again a C$^*$--algebra but, in contrast 
to $\bpi (\fA)$, it is also closed with respect to weak limits. 
Such weakly closed algebras are called 
von Neumann algebras.\footnote{Similarly to the 
case of C$^*$--algebras, there exists also an abstract version of 
von Neumann algebras, the W$^*$--algebras. In the present context, 
we do not need to distinguish between those.}

In the analysis of von Neumann algebras ${\cal R}$ one uses 
various notions which enter also in the present discussion. 
The ``center'' $\cal Z$ of $\cal R$ is the subalgebra 
of operators commuting with all operators in $\cal R$.
If this center consists only of multiples of the unit operator, $\cal
R$ is called a ``factor''. The decomposition of an algebra into
factors is unique and corresponds to the simultaneous 
spectral resolution of all operators in the
center. Another important notion is ``hyperfinite'', which
means that the algebra can be approximated (in the sense of 
weak limits) by its finite dimensional subalgebras. 

In their seminal investigation, entitled ``rings of operators'', 
von Neumann and Murray found that there were several types of
factors. The ones with which physicists were familiar (``type I'')
correspond to the algebra of all bounded operators on some  
Hilbert space. Different factors of type I can thus be distinguished 
by the dimension of the underlying space. 
Then there was a continuous generalization, called type II, in
which a trace could still be defined for a class of
elements. Everything else was lumped together as ``type III''. 

The Tomita--Takesaki theory provided tools for a finer subdivision. 
In this theory, the basic ingredients are, besides a von
Neumann algebra $\cal{R}$, cyclic and separating vectors $\Psi$,
i.e.\ vectors for which $\cal R\Psi$ is dense in the
Hilbert space and which are annihilated by
none of the operators in $\cal R$, apart from $0$. Given 
such a pair $({\cal R}, \Psi)$, one can consistently define an antilinear
operator $S$, the Tomita conjugation, setting
\begin{equation}
S \, A \Psi = A^* \Psi \quad \mbox{for} \ A \in {\cal R}.
\end{equation}
The conjugation $S$ can be decomposed in a unique way 
(polar decomposition) into the   
product $S = J \, \Delta^{1/2}$ of an anti--unitary operator 
$J$ and a positive self--adjoint operator $\Delta^{1/2}$
whose square is called the modular operator affiliated with
$({\cal R}, \Psi)$. It is a central result in this theory that 
the corresponding unitary operators $\Delta^{it}$, $t \in \RR$, map
by their adjoint action the algebra ${\cal R}$ onto itself.
These maps are 
the modular automorphisms mentioned at various points in the main text. 

Based on these notions, an
essentially complete classification of  
factors ${\cal R}$ was achieved by Alain Connes  
\cite{Con}, who showed that the ``spectral invariant'', obtained from the
intersection of the spectra of modular operators
affiliated with ${\cal R}$, is (disregarding the value 0) 
always a closed subgroup of the multiplicative group of positive
real numbers. All such groups    
occur in this classification, but we mention
only the ones which will concern us here. In the case of the
types I and II the group consists only of the unit
element. The opposite situation is
that the group consists of all positive real numbers. This was called
``type III$_1$''. As was shown later by Haagerup \cite{Haa}, 
the hyperfinite factor of type III$_1$ is unique.
It is this factor which generically appears in quantum 
field theory.\newpage 
{\noindent \Large \bf Acknowledgment} \\[3mm]
We are grateful to John E.\ Roberts for several useful suggestions.
%%%%%%%%%%%%%%%%%%%%%%%%%%%%%%%%%%%%%%%%%%%%%%%%%%%%%%%%%%%%%%%%%%%%%%
%%%%%%%%%%%%%%%%%%%%%%%%%%%%%%%%%%%%%%%%%%%%%%%%%%%%%%%%%%%%%%%%%%%%%%


\begin{thebibliography}{99}
\small 
\bibitem{StWi}
R.F.\ Streater and A.S.\ Wightman,
{\em PCT, Spin and Statistics, and all that},  
Benjamin 1964 

\bibitem{Jo} 
R.\ Jost,  
{\em The General Theory of Quantized Fields}, 
American Math. Soc. 1965 

\bibitem{BoSh} N.N.\ Bogolubov and D.V.\ Shirkov, 
{\em Introduction to the theory of quantized fields}, 
Interscience 1958 

\bibitem{BoLoTo} 
N.N.\ Bogolubov, A.A.\ Logunov and I.T.\ Todorov, 
{\em Introduction to Axiomatic Quantum Field Theory}, 
Benjamin 1975

\bibitem{Ha}
R.\ Haag, 
{\em Local Quantum Physics: Fields, Particles, Algebras},
2$^{\rm nd}$ revised ed.,  
Springer 1996 

\bibitem{LeSyZi}
H.\ Lehmann, K.\ Symanzik and W.\ Zimmermann, 
Nuovo Cimento {\bf 1} (1955) 425

\bibitem{Wig} A.S.\ Wightman, Phys.\ Rev.\ {\bf 101} (1956) 860

\bibitem{Ek} H.\ Ekstein, Phys.\ Rev.\ {\bf 101} (1956) 880
\ and \ Nuovo Cimento {\bf 4} (1956) 1017 

\bibitem{Ha1} R.\ Haag, Phys.\ Rev.\ {\bf 112} (1958) 669

\bibitem{BrHa} W.\ Brenig and R.\ Haag, Fortschr.\ Phys.\ {\bf 7}
(1959) 183

\bibitem{Ru} D.\ Ruelle, Helv.\ Phys.\ Acta {\bf 35} (1962) 147

\bibitem{St} F.\ Strocchi, 
p.\ 551 in: {\em Mathematical Methods in 
Theoretical Physics}, Proceedings Boulder Col.\ 1971, W.E.Brittin ed., 
Colorado Assoc.\ Univ. Press 1973

\bibitem{KuOj} T.\ Kugo and I. Ojima, 
Prog.\ Theor.\ Phys.\ Suppl.\ {\bf 66} (1979) 1

\bibitem{KaRi} R.V.\ Kadison and J.R.\ Ringrose,
{\it Fundamentals of the Theory of Operator Algebras, Vol.\ 1 and 
Vol.\ 2}, Academic Press 1986

\bibitem{BrRo} O.\ Bratteli and D.W.\ Robinson, {\it Operator Algebras
and Quantum Statistical Mechanics, Vol.\ 1}, Springer 1979

\bibitem{ArHa} H.\ Araki and R.\ Haag, Commun.\ Math.\ Phys.\ {\bf 4}
(1967) 77

\bibitem{BuPoSt} D.\ Buchholz, M.\ Porrmann and U.\ Stein, 
Phys.\ Lett.\ {\bf B267} (1991) 377 

\bibitem{Bo} H.-J.\ Borchers, Nuovo Cimento {\bf 15} (1960) 784 

\bibitem{Re} K.--H.\ Rehren,
Commun.\ Math.\ Phys. {\bf 178} (1996) 453 

\bibitem{Co} S.\ Coleman, 
Phys.\ Rev.\ {\bf D11} (1975) 2088

\bibitem{Se} 
K.\ Intriligator and N.\ Seiberg, 
Nucl.\ Phys.\ Proc.\ Suppl.\ {\bf 45BC} (1996) 1 

\bibitem{BoZi} H.--J.\ Borchers and W.\ Zimmermann,
Nuovo Cimento {\bf 31} (1963) 1047

\bibitem{DrFr}
W.\ Driessler and J.\ Fr\"ohlich, 
Ann.\ Inst.\ H.\ Poincar\'e {\bf 27} (1977) 221

\bibitem{BoYn}
H.--J.\ Borchers and J.\ Yngvason,
Rev.\ Math.\ Phys.\ {\bf Special Issue} (1992) 15

\bibitem{FrHe}
K.\ Fredenhagen and J.\ Hertel,
Commun.\ Math.\ Phys.\ {\bf 80} (1981) 555

\bibitem{HaOj}
R.\ Haag and I.\ Ojima,
Ann.\ Inst.\ H.\ Poincar\'e {\bf 64} (1996) 385

\bibitem{WiWiWi} J.C.\ Wick, A.S.\ Wightman and E.P.\ Wigner, 
Phys.\ Rev.\ {\bf 88} (1952) 101

\bibitem{HaKa} R.\ Haag and D.\ Kastler, J.\ Math.\ Phys.\ {\bf 5}
(1964) 848

\bibitem{HaHuWi} R.\ Haag, N.M.\ Hugenholtz and M.\ Winnink, 
Comm.\ Math.\ Phys.\  {\bf 5} (1967) 215

\bibitem{Ta} M.\ Takesaki, {\it Tomita's Theory of Modular
Hilbert Algebras and its Applications}, Lecture Notes in Mathematics, 
Springer 1970 

\bibitem{ReSch} H.\ Reeh and S.\ Schlieder, Nuovo Cimento {\bf 22}
(1961) 1051 

\bibitem{BiWi} J. Bisognano and E.H. Wichmann, 
J.\ Math.\ Phys.\ {\bf 16} (1975) 985 \ and \ 
J.\ Math.\ Phys.\ {\bf 17} (1976) 303

\bibitem{Sew} G. Sewell, 
Phys.\ Rev.\ Lett.\ {\bf 79A} (1980) 23

\bibitem{HaNaSt} R.\ Haag, H.\ Narnhofer and U.\ Stein, 
Commun.\ Math.\ Phys.\ {\bf 94} (1984) 219  

\bibitem{HiLo} P.\ Hislop and R.\ Longo, Commun.\ Math.\ Phys.\
{\bf 84} (1982) 71

\bibitem{HaSw} R.\ Haag and J.A.\ Swieca, 
Commun.\ Math.\ Phys.\ {\bf 1} (1965) 308

\bibitem{BuWi}
D.\ Buchholz and E.H.\ Wichmann, 
Commun.\ Math.\ Phys.\ {\bf 106} (1986) 321

\bibitem{BuJu} D.\ Buchholz and P.\ Junglas, 
Commun.\ Math.\ Phys.\ {\bf 121} (1989) 255

\bibitem{BrBu} J.\ Bros and D.\ Buchholz,
Nucl.\ Phys.\ {\bf B429} (1994) 291, \
Ann.\ Inst.\ H.\ Poincar\'e {\bf 64} (1996) 495, \ and \ 
Z.\ Phys.\ {\bf C55} (1992) 509 

\bibitem{Fr} K.\ Fredenhagen, Commun.\ Math.\ Phys.\ {\bf 97} (1985) 79

\bibitem{BuDaFr}
D.\ Buchholz, C.\ D'Antoni and K.\ Fredenhagen, 
Commun.\ Math.\ Phys.\ {\bf 111} (1987) 123

\bibitem{Haa} U.\ Haagerup, 
Acta Math. {\bf 158} (1987) 95

\bibitem{SuWe}
S.J.\ Summers and R.\ Werner, Commun.\ Math.\ Phys.\ {\bf 110} (1987)
1004 \ and \ Lett.\ Math.\ Phys.\ {\bf 33} (1995) 321

\bibitem{DoLo} S.\ Doplicher and R.\ Longo, Invent.\ Math.\ {\bf 73}
(1984) 493

\bibitem{Gr} H.S.\ Green, Phys.\ Rev.\ {\bf 90} (1953) 270

\bibitem{BaWo} H.\ Baumg\"artel and M.\ Wollenberg,
{\it Causal Nets of Operator Algebras}, 
Akademie Verlag 1992

\bibitem{Ka} D.\ Kastler (ed.), 
{\it The Algebraic Theory of Superselection Sectors and Field Theory:
Introduction and Recent Results}, Proceedings Palermo 1989,  
World Scientific 1990
 
\bibitem{DoHaRo} S.\ Doplicher, R.\ Haag and J.E.\ Roberts, 
Commun.\ Math.\  Phys.\ {\bf 23} (1971) 199 \ and \ 
Commun.\ Math.\  Phys.\ {\bf 35} (1974) 49

\bibitem{Sw} J.A.\ Swieca, 
Phys.\ Rev.\ {\bf D13} (1976) 312

\bibitem{DoRo} S.\ Doplicher and J.E.\ Roberts,  
Commun.\ Math.\ Phys.\ {\bf 131} (1990) 51

\bibitem{FrReSch} K.\ Fredenhagen, K.-H.\ Rehren and B.\ Schroer, 
Commun.\ Math.\ Phys.\ {\bf 125} (1989) 201 \ and \ 
Rev.\ Math.\ Phys.\ {\bf Special Issue} (1992) 111  

\bibitem{Fro} 
J.\ Fr\"ohlich and F. Gabbiani,  
Rev.\ Math.\ Phys.\ {\bf 2} (1991) 251

\bibitem{Sch1}
B.\ Schroer, {\em Modular Localization and Nonperturbative Local
Quantum Physics}, Lecture Notes, Rio de Janeiro: CBPF, 1998,
preprint hep-th/9805093

\bibitem{BuFr} D.\ Buchholz and K.\ Fredenhagen, 
Commun.\ Math.\ Phys.\ {\bf 84} (1982) 1 

\bibitem{BrEp} J.\ Bros and H.\ Epstein, 
p. 330 in: {\em XIth International Congress of Mathematical
Physics. Paris 1994\/}, D.\ Iagolnitzer ed., Internat.\ Press 1995

\bibitem{Bu2} D.\ Buchholz, Commun.\ Math.\ Phys.\, 
{\bf 85} (1982) 85                                 

\bibitem{Sch} B.\ Schroer, Fortschr.\ Phys.\ {\bf 11} (1963) 1

\bibitem{FrMoSt} J.\ Fr\"ohlich, G.\ Morchio and F.\ Strocchi, 
Annals Phys.\ {\bf 119} (1979) 241

\bibitem{Bu3} D.\ Buchholz,
Phys.\ Lett.\ {\bf B174} (1986) 331

\bibitem{Ma} G.\ Mack, 
Fortsch.\ Phys.\ {\bf 29} (1981) 135

\bibitem{Ar} H.\ Araki, 
Rev.\ Math.\ Phys.\  {\bf Special Issue} (1992) 1 

\bibitem{BuDoLo} D.\ Buchholz, S.\ Doplicher and R.\ Longo,
Ann.\ Phys.\ {\bf 170} (1986) 1 

\bibitem{Sei} E.\ Seiler, 
{\it Gauge Theories as a Problem of Constructive Quantum Field Theory
and Statistical Mechanics}, Springer  1982

\bibitem{Ste} O.\ Steinmann, 
Ann.\ Phys.\ {\bf 157} (1984) 232  

\bibitem{BuVe}
D.\ Buchholz and R.\ Verch,
Rev.\ Math.\ Phys.\ {\bf 7} (1995) 1195 \ and \
Rev.\ Math.\ Phys.\ {\bf 10} (1998) 775

\bibitem{WiZi}
K.\ Wilson and W.\ Zimmermann, 
Commun.\ Math.\ Phys.\ {\bf 24} (1971) 87

\bibitem{Bu4} 
D.\ Buchholz,
Nucl.\ Phys.\ {\bf B469} (1996) 333

\bibitem{GlJa} J.\ Glimm and A.\ Jaffe, {\it Quantum Physics: A 
Functional Integral Point of View}, Springer 1987 


\bibitem{Lo} R.\ Longo,
Commun.\ Math.\ Phys.\ {\bf 126} (1989) 217 \ and \
Commun.\ Math.\ Phys.\ {\bf 130} (1990) 285

\bibitem{Wi} H.W.\ Wiesbrock,
Commun.\ Math.\ Phys.\ {\bf 157} (1993) 83, Erratum 
Commun.\ Math.\ Phys.\ {\bf 184} (1997) 683, \ and \ 
Commun.\ Math.\ Phys.\ {\bf 193} (1998) 269

\bibitem{KaWi} R.\ K\"ahler and H.W.\ Wiesbrock, 
``Modular theory and the reconstruction of 4--dimensional 
quantum field theories'', preprint   

\bibitem{Bo2} H.-J.\ Borchers, 
Commun.\ Math.\ Phys.\ {\bf 143} (1992) 315
 
\bibitem{BuDrFlSu} 
D.\ Buchholz, O.\ Dreyer, M.\ Florig and S.J.\ Summers, 
``Geometric modular action and spacetime symmetry groups'',
preprint math-ph/9805026, to appear in Rev.\ Math.\ Phys. 

\bibitem{Di}  J.\ Dimock, Commun.\ Math.\ Phys.\ {\bf 77} (1980) 219 

\bibitem{Wa} R.\ Wald, 
{\it Quantum Field Theory in Curved Spacetime and Black Hole
Thermodynamics},  Univ.\ Chicago Pr.\ 1994

\bibitem{BrFrKo}
R.\ Brunetti, K.\ Fredenhagen and M.\ K\"ohler, 
Commun.\ Math.\ Phys.\ {\bf 180} (1996) 633

\bibitem{Ha2} R.\ Haag, Commun.\ Math.\ Phys.\ {\bf  123} (1990) 245, \
\ Commun.\ Math.\ Phys.\ {\bf 180} (1996) 733, \ 
Z.\ Naturforschung {\bf 54a} (1999) 2, \ and \ 
``Objects, Events, Localization'', 
ESI--preprint 541 (1998)

\bibitem{Con} A.\ Connes, 
Ann.\ Sci.\ Ecole Norm.\ Sup.\ {\bf 6} (1973) 133


%%%%%%%%%%%%%%%%%%%%%%%%%%%%%%%%%%%%%%%%%%%%%%%%%%%%%%%%%%%%%%%%%%%%%%%
%%%%%%%%%%%%%%%%%%%%%%%%%%%%%%%%%%%%%%%%%%%%%%%%%%%%%%%%%%%%%%%%%%%%%%%
%%%%%%%%%%%%%%%%%%%%%%%%%%%%%%%%%%%%%%%%%%%%%%%%%%%%%%%%%%%%%%%%%%%%%%%
\end{thebibliography}
\end{document}